\documentclass[intlimits,twoside,a4paper]{article}

\usepackage[cp1251]{inputenc}

\usepackage[eqsecnum]{cmpj3}

\usepackage{graphicx}

\articletype{A part of the Special collection to the 100th anniversary of the birth of Ihor Yukhnovskii}

\issue{2025}{28}{2}{23602}
\doinumber{10.5488/CMP.28.23602}

\title[Replica RISM for nanoporous materials]%
{Replica RISM molecular solvation theory for electric double layer in nanoporous materials}

\author[A. Kovalenko]{A. Kovalenko\orcid{0000-0001-5033-4314}\thanks{Corresponding author: \email{akovalenko@smmodeling.com}}}

\address{Software for Multiscale Modeling Inc., Edmonton, Alberta, Canada T6E 5J5}


\date{Received March 18, 2025, in final form April 3, 2025}
\begin{document}

\maketitle

\begin{abstract}

Applications of 3D-RISM-KH molecular solvation theory range from solvation energy of small molecules to phase behavior of polymers and biomolecules. It predicts the molecular mechanisms of chemical and biomolecular systems. Replica RISM-KH-VM molecular solvation theory predicts and explains the structure, thermodynamics, and electrochemistry of electrolyte solutions sorbed in a nanoporous material. It was tested on nanoporous carbon supercapacitors with aqueous electrolyte and nanoporous electrosorption cells. The mechanisms in these systems are steered by the electric double layer potential drop across the Stern layer at the nanopores surface and the Gouy-Chapman layer averaged over the nanoporous material, the osmotic term due to the ionic concentrations difference in the two nanoporous electrodes and in the electrolyte solution outside, and the solvation chemical potentials of sorbed ions averaged over the nanoporous material. The latter strongly depends on chemical specificity of ions, solvent, surface functional groups, and steric effects for solvated ions confined in nanopores.

\printkeywords

\end{abstract}

\setcounter{equation}{0}

\section{Introduction}

Molecular properties are considerably different from continuous media because of various scalability of these systems. The driving factors in such behaviors stem from the different forms due to size, shape, composition, and physical states. Understanding all the interactions between different constituent fragments is critical to build materials with desired properties. It is difficult to attain this goal from experimental point of view due to numerous possible combinations between structure and the resultant reactivity and activity. Theoretical prediction can extend over the entire range of the size scale and to the larger part of the time scale. Predictive modelling capabilities can address the issues related to single molecules in the realm of complex molecules and nanomaterials. Predictive modelling for sophisticated electronic structure methods is incompatible with molecular dynamics simulations for real systems with millions or billions of atoms, since the computing cost becomes astronomical. A multiscale predictive method should possess a sufficient capability to couple lower and upper scale methods by propagating the accuracy of the lower scale methods to the higher scale systems without extensive computational requirements. The gas-phase chemical physics level is mostly impossible to be applied to realistic systems not only due to the size but also due to the presence of a solvent.

Quantum mechanical calculations for systems in continuum media using continuum solvation models are useful if the systems have a reasonable size to be treated with at least a triple-$\zeta$ basis set. Unfortunately, the presence of heavy elements makes this approach useless. The applicability of the Onsager equation in predicting the solvation free energy was questioned \cite{Zhang:2017}. Possible schemes in handing (bio)chemical processes in a solution are classified either as those using continuum solvation models, the so-called cluster continuum models or quasi-chemical models, all having the system size limitations \cite{Gerez:2023, Skyner:2015, Vyboishchikov:2021, Klamt:2015, Chaudhari:2020, Pratt:1999, Asthagiri:2010, Rogers:2011, Rogers:2012}. The other area is based on the statistical mechanics, or the reference interaction site model (RISM), that acquired popularity in treating molecular systems in liquids due to a reasonable accuracy, calculation speed, and extendibility over the entire scale of the molecular size and time scales \cite{Hirata:2003, Roy:2021}. Molecular solvation theory with integral equation formalism is based on the modified Ornstein-Zernike (OZ) theory for dimensional reduction of molecular liquids \cite{Hansen:2013}. The key to the RISM theory are the spatial distributions and statistical mechanical ensembles in order to predict the solvation thermodynamics via integration over an infinite number of interaction diagrams. Its three-dimensional version, 3D-RISM theory \cite{Kovalenko:1998, Kovalenko:1999, Kovalenko:2000-1, Kovalenko:2000-2, Kovalenko:2003}, gives the 3D maps of solvent distributions around a solute (macro)molecule. A successful implementation of the 3D-RISM integral equation needs a closure relation, which can be generalized as a functional to impose a consistency condition of the path independent chemical potential $\mu$. During over forty years of RISM theory, only a handful of closure relations exists, depending on the accuracy and scope of applications \cite{Chandler:1974, Lowden:1973, Chandler:1973}. The most promising closure relation is the Kovalenko-Hirata (KH) closure approximation \cite{Kovalenko:1999} that was shown to work for small molecules, biomolecules, nanomaterials, aggregates, and ligand-protein interactions, with high accuracy in the whole range of thermodynamic conditions \cite{Johnson:2016, Kondratenko:2015, Genheden:2014, Stoyanov:2013, Stoyanov:2008, Imai:2005}.

The replica RISM-KH-VM (modified Verlet) theory for electrolyte solutions sorbed in disordered nanoporous materials stems from the replica formalism in statistical mechanics of quenched-annealed systems. Replica RISM-KH-VM produces the solvation structure, thermochemical, and electrochemistry averaged over the thermal motion of sorbed solution as well as the quenched morphology and functionalization of host nanoporous material. It predicts the solvent-specific wetting and water depletion in hydrophobic nanopores, asymmetry in solvation and adsorption of cations and anions, desalination of simple ions in hydrophobic nanopores, desalination reversal with external voltage, and specific adsorption in functionalized carbon nanopores. Replica RISM-KH-VM reveals that the mechanisms of a nanoporous supercapacitor and nanoporous electrosorption cell are much more complex than the common view of a planar electrical double layer mapped onto the nanoporous morphology, and are determined by several factors in the Nernst equation. The chemical potentials consist of the ideal term for the different concentrations of solution species in the nanoporous cathode and anode, and in the bulk electrolyte solution outside (i.e., osmotic effect). The solvation free energy of sorbed electrolyte solution is modified by nanoporous confinement and effective interaction with the bulk of functionalized nanoporous material statistically averaged over the nanoporous morphology (nanoporous material volume effect). The electric potential step of Gibbs-Donnan type across a diffuse double layer at the nanoporous electrode boundary in contact with the bulk solution keeps the chemical balance (i.e., electrode boundary effect). The voltage between the charged nanoporous cathode and anode electrodes and the bulk solution outside is determined by the electrostatic potential change from the surface of carbon nanopores to the mean potential level inside the nanoporous electrode (Stern surface layer and Gouy-Chapman diffuse layer), plus the Gibbs-Donnan electric potential step from the nanoporous electrode volume to the solution bulk. This picture generalizes the traditional description of nanoporous electrodes based on an equivalent planar electric double layer and Donnan potential to a molecular description with coupled electrostatics, molecular specificity of solvent and electrolyte, adsorption at functionalized nanopores, accommodation of ions with their solvation shells in nanopores, and osmotic effects.

\section{DRISM-KH theory for ion-molecular liquid mixtures}

The RISM molecular solvation theory relies on the probability density $\rho_\alpha g_{\alpha\gamma}\left(r\right)$ of finding interaction sites $\alpha$ and $\gamma$ of solvent molecules at distance $r$, with the average solvent site number density $\rho_\alpha$ in the solution bulk and the normalized solvent site density distribution function $g_{\alpha\gamma}\left(r\right)$. The solvent site-site total correlation function $h_{\alpha\gamma}\left(r\right) = g_{\alpha\gamma}\left(r\right) - 1$ is obtained from the dielectrically consistent reference interaction site model (DRISM) theory \cite{Perkyns:1992} complemented with the Kovalenko-Hirata (KH) closure approximation \cite{Kovalenko:1999}. DRISM theory enforces the correct asymptotics of the correlation functions and the given dielectric constant in a mixture of polar solvent with ions at a finite concentration. It was applied to the bulk solution of a given composition, including polar solvent, co-solvent, electrolyte, and ligands at a given concentration. The DRISM integral equation for the site-site total $h_{\alpha\gamma}\left(r\right)$ and direct $c_{\alpha\gamma}\left(r\right)$ correlation functions of solvent has the form \cite{Perkyns:1992}
\begin{equation}
  \tilde{h}_{\alpha\gamma}\left(r\right)
  = \tilde{\omega}_{\alpha\mu}\left(r\right) * c_{\mu\nu}\left(r\right) * \tilde{\omega}_{\nu\gamma}\left(r\right) 
    + \tilde{\omega}_{\alpha\mu}\left(r\right) * c_{\mu\nu}\left(r\right) * \rho_\gamma \tilde{h}_{\nu\gamma}\left(r\right) .
\label{eqn:drism}
\end{equation}
Both the intramolecular site-site correlation function $\tilde{\omega}_{\alpha\mu}\left(r\right)$ and the site-site total correlation function $\tilde{h}_{\alpha\mu}\left(r\right)$ in the DRISM equation (\ref{eqn:drism}) are renormalized with an analytical dielectric bridge correction enforcing all the correct inter- and intra-species interactions, 
\begin{equation}
  \tilde{\omega}_{\alpha\gamma}\left(r\right)	= \omega_{\alpha\gamma}\left(r\right) + \rho_\alpha \chi_{\alpha\gamma}\left(r\right) ,
  \label{eqn:drism-renorm-omega}
\end{equation}
\begin{equation}
  \tilde{h}_{\alpha\gamma}\left(r\right) = h_{\alpha\gamma}\left(r\right) - \chi_{\alpha\gamma}\left(r\right) .
  \label{eqn:drism-renorm-h}
\end{equation}
The intramolecular correlation function $\omega_{\alpha\gamma}\left(r\right) = \delta\left(r-l_{\alpha\gamma}\right)$ represents the geometry of rigid solvent molecules with site-site separations $l_{\alpha\gamma}$. It is specified in the reciprocal k-space in terms of the zeroth-order spherical Bessel function $j_0(x)$ as
\begin{equation}
  \omega_{\alpha\gamma}\left(k\right) = j_0\left(kl_{\alpha\gamma}\right) .
  \label{eqn:omega-intra}
\end{equation}
The term $\chi_{\alpha\gamma}$ in the renormalized dielectric correction has the following form in the reciprocal $k$-space~\cite{Perkyns:1992}:
\begin{equation}
  \chi_{\alpha\gamma}(k) = j_0\left(kx_\alpha\right) j_0\left(ky_\alpha\right) j_1\left(kz_\alpha\right) j_0\left(kx_\gamma\right) j_0\left(ky_\gamma\right) j_1\left(kz_\gamma\right) ,
  \label{eqn:chi-drism}
\end{equation}
where $j_0(x)$ and $j_1(x)$ are the zeroth- and first-order spherical Bessel functions over the positions of each atom ${\bf r}_\alpha = \left(x_\alpha,y_\alpha,z_\alpha\right)$ with partial site charge $q_\alpha$ of site $\alpha$ on species $s$ with respect to its molecular origin, where both sites $\alpha$ and $\gamma$ are on the same species $s$, and its dipole moment ${\bf d}_s = \sum_{\alpha \in s} q_\alpha {\bf r}_\alpha$ is oriented along the $z$-axis, ${\bf d}_s = \left(0,0,d_s\right)$. The renormalized dielectric correction (\ref{eqn:chi-drism}) is nonzero only for polar solvent species of the sorbed electrolyte solution which possess a dipole moment and are responsible for the dielectric response in the DRISM approach. The value of the envelope function $h_{\rm c}(k)$ at $k=0$ determines the dielectric constant of the solution, and is assumed in a smooth non-oscillatory form quickly falling off at wavevectors $k$ larger than those corresponding to a characteristic size $l$ of liquid molecules and hence to the dielectric constant $\varepsilon$ of the solvent,
\begin{equation}
  h_{\rm c}\left(k\right) = A \exp \left( -l^2 k^2 / 4 \right) ,
  \label{eqn:hc}
\end{equation}
and 
\begin{equation}
  A = \frac{1}{\rho_{\rm polar}} \left( \frac{\varepsilon}{y} - 3 \right) ,
  \label{eqn:A}
\end{equation}
where $A$ is the amplitude. In the dielectric correction form (\ref{eqn:chi-drism})--(\ref{eqn:A}), mixed solvents have the total number density of solution polar species
\begin{equation}
  \rho_{\rm polar} = \sum_{s \in {\rm polar}} \rho_s 
  \label{eqn:rho-polar}
\end{equation}
and the dielectric susceptibility
\begin{equation}
  y = \frac{4\piup}{9k_\textrm{B}T} \sum_{s \in {\rm polar}} \rho_s \left(d_s\right)^2 .
  \label{eqn:diel-susc}
\end{equation}
The parameter $l$ in the form (\ref{eqn:hc}) specifies the characteristic separation below which the correction is turned off so as not to distort the short-range solvation structure. It should be $l=1$~\AA~for water, and larger for solvents with larger molecules.

A closure relation between the site-site total and direct correlation functions that complements the DRISM integral equations~(\ref{eqn:chi-drism}) provides a computational approach to integrate the infinite chain of diagrams. The exact closure can be formally expressed as a series in terms of multiple integrals of the combinations of the total correlation functions. However, it is extremely cumbersome, and is replaced with practical approximations. Among a number of closure relations, the Kovalenko-Hirata (KH) closure was shown to work for small molecules, biomolecules, nanomaterials, aggregates, ligand-protein interactions, with high accuracy in the whole range of thermodynamic conditions \cite{Johnson:2016, Kondratenko:2015, Genheden:2014, Stoyanov:2013, Stoyanov:2008, Imai:2005}. It consistently accounts for both electrostatic, associative, and steric effects of solvation in simple and complex liquids. The KH closure relation to the DRISM integral equations is written as
\begin{equation}
  g_{\alpha\gamma}(r) = \begin{cases}
     \exp\left( -u_{\alpha\gamma}(r) / \left(k_\textrm{B}T\right) + h_{\alpha\gamma}(r) - c_{\alpha\gamma}(r) \right) & \text{for } g_{\alpha\gamma}(r) \leqslant 1 \\
     1 - u_{\alpha\gamma}(r) / \left(k_\textrm{B}T\right) + h_{\alpha\gamma}(r) - c_{\alpha\gamma}(r) & \text{for } g_{\alpha\gamma}(r) > 1 
	 \end{cases} ,
  \label{eqn:KH-closure}
\end{equation}
where $u_{\alpha\gamma}(r)$ is the interaction potential between the solvent sites $\alpha$ and $\gamma$ specified by the molecular force field. The KH closure (\ref{eqn:KH-closure}) nontrivially couples the mean spherical approximation (MSA) for spatial regions of solvent density enrichment $_{\alpha\gamma}(r) > 1$ and the hypernetted chain (HNC) one for density depletion $g_{\alpha\gamma}(r) \leqslant 1$. The site-site distribution function and its first derivative are continuous at the joint boundary $g_{\alpha\gamma}(r) = 1$ by construct. The RISM-HNC theory overestimates attraction in strongly associated solvents, thus imparting numerical instability, and diverges for strongly charged systems such as electrolytes. The DRISM-KH theory easily handles all such systems. The other closure relations that developed over the years are aimed at specific applications and lack the generality of the KH closure~\cite{Percus:1958, Martynov:1983, Ballone:1986, Kast:2008}. The KH closure underestimates the height of strong associative peaks although it widens the peaks and provides correct thermodynamics and solvation structure \cite{Perkyns:2011, Stumpe:2011}.

The DRISM-KH integral equations have an exact differential of the solvation free energy that yields an analytical expression of Kirkwood’s thermodynamic integration gradually switching on the solute-solvent interaction. The solvation free energy of a molecule in multicomponent solvent is thus obtained in a closed analytical form as a single integral of the solvent site-site correlation functions:
\begin{equation}
  \mu_{\rm solv} = 4\piup \sum_{\alpha\gamma} \int_0^\infty  {\rm d}r \,r^2 \Phi_{\alpha\gamma}(r) ,
  \label{eqn:SFE}
\end{equation}
\begin{equation}
  \Phi_{\alpha\gamma}(r) = \rho_\gamma k_{\rm B}T \left( \frac{1}{2} h^2_{\alpha\gamma}(r)\Theta\left(-h_{\alpha\gamma}(r)\right) - c_{\alpha\gamma}(r) - \frac{1}{2} h_{\alpha\gamma}(r) c_{\alpha\gamma}(r) \right) ,
  \label{eqn:SFED}
\end{equation}
where $\Theta(x)$ is the Heaviside step function. The integrand $\Phi_{\alpha\gamma}(r)$ in equation (\ref{eqn:SFE}) is the solvation free energy density arising due to all the solvent-solvent interactions. The solvation free energy of a solute molecule $\Delta\mu$ is obtained by summation of the partial contributions over all solvent sites integrated over the whole volume. Other thermodynamics quantities are derived from the solvation free energy (\ref{eqn:SFE}) by differentiation. This includes the solvation chemical potential which is decomposed into the energetic and entropic components at constant volume,
\begin{equation}
  \Delta\mu = \Delta\varepsilon^{\rm uv} = \Delta\varepsilon^{\rm vv} - T\Delta s_{\rm V} ,
  \label{eqn:SFE-decomp}
\end{equation}
where entropy at a constant volume is
\begin{equation}
  \Delta s_{\rm V} = - \frac{1}{T} \left( \frac{\partial\Delta\mu}{\partial T} \right)_{\rm V} ,
  \label{eqn:entropy}
\end{equation}
the internal energy of the solute-solvent (``uv'') interaction is
\begin{equation}
  \Delta\varepsilon^{\rm uv} = k_{\rm B}T \sum_{\alpha\gamma} \rho_\alpha 4\piup \int_0^\infty {\rm d}r\,r^2  g_{\alpha\gamma}(r) u_{\alpha\gamma}(r) ,
  \label{eqn:energy-uv}
\end{equation}
and the remaining term $\Delta\varepsilon^{\rm vv}$ gives the energy of solvent reorganization around the solute. The partial molar volume of the solute macromolecule is obtained from the Kirkwood-Buff theory \cite{Kirkwood:1951}  extended to the RISM formalism as \cite{Harano:2001, Imai:2001}
\begin{equation}
  V = k_{\rm B}T\chi_T \left( 1 - 4\piup \sum_{\alpha\gamma} \rho_\alpha \int_0^\infty {\rm d}r\,r^2  c_{\alpha\gamma}(r) \right) ,
  \label{eqn:PMV}
\end{equation}
where $\chi_T$ is the isothermal compressibility of the bulk solvent obtained in terms of the site-site direct correlation functions of bulk solvent as 
\begin{equation}
  \rho k_{\rm B}T\chi_T = \left( 1 - 4\piup \sum_{\alpha\gamma} \rho_\alpha \int_0^\infty {\rm d}r\,r^2  c_{\alpha\gamma}(r) \right)^{-1} ,
  \label{eqn:chi}
\end{equation}
and $\rho=\sum_s \rho_s$ is the total number density of the bulk solvent mixture of molecular species $s$. To apply DRISM-KH theory in calculating the absolute solvation free energy, one should be careful as the theory overestimates the solvation by a large margin, although it produces correct trends. To avoid overestimation, one can use the so-called universal correction (UC) scheme \cite{Palmer:2010}
\begin{equation}
  \Delta\mu_{\rm UC} = \Delta\mu_{\rm GF} + aV + b .
  \label{eqn:UC}
\end{equation}
The chemical potential adjusted with the UC is obtained from the Gaussian Fluctuation chemical potential $\Delta\mu_{\rm GF}$ corrected with partial molar volume $V$. The correlation coefficients a and b in this expression are obtained from multiple linear regression analysis carried out against benchmarking results. Such applications were reported, too \cite{Luchko:2016, Roy:2017}.

The most efficient way of converging the DRISM-KH equation is by using the modified direct inversion in the iterative subspace (MDIIS) accelerated numerical solver \cite{Kovalenko:1999-MDIIS}. This is a simple and fast protocol with much smaller computer memory demand to solve for the direct correlation functions. It zeroes the residuals which are nonlocal functionals of the direct correlation functions and are computed as a difference between the site-site distribution functions generated from the integral equation (\ref{eqn:drism}) and the closure relation (\ref{eqn:KH-closure}). MDIIS is simple, robust and stable, has relatively small memory usage, and provides a substantial acceleration with quasiquadratic convergence in the whole range of the root mean square residual. It reliably converges for complex charged systems with strong associative and steric effects, which is challenging for the integral equations. MDIIS is closely related to Pulay’s DIIS solver to accelerate and stabilize the convergence of the Hartree–Fock self-consistent field equations \cite{Pulay:1980}. Other similar algorithms include the generalized minimal residual (GMRes) solver \cite{Saad:1986} which was also coupled with a Newton-Raphson-like approach on a coarse grid \cite{Howard:2008} as well as its limitation to the solute repulsive core~\cite{Minezawa:2007}.

\section{Replica RISM theory for electrical double layer in nanoporous materials}

\setcounter{equation}{18}

Due to the overlap of electric double layers (EDLs) in nanoporous carbon electrodes, they are different from planar electrochemical capacitors. Unlike a planar electrode, the EDL at the inner surface of nanopores is distorted and has a specific capacitance higher by 1-2 orders of magnitude. Another EDL at the outer macroscopic surface of nanoporous electrodes strongly contributes to the specific capacitance. Molecular simulation of these systems is practically unfeasible due to the interplay of long-range electrostatic and short-range interactions as well as the chemical and mechanical balance between the sorbed electrolyte solution and that in the bulk. Conventional modelling employed MD simulation for an EDL with finite size ions in slit-like pores with no molecular solvent and chemical specificities~\cite{Skinner:2011} and all-atom MD simulation of ions and solvent in confinement with simplified geometry \cite{Feng:2012}.

The replica RISM-KH-VM theory was developed by generalizing RISM molecular solvation theory to solutions sorbed in a disordered nanoporous material \cite{Kovalenko:2002, Kovalenko:2001-1, Kovalenko:2001-2, Kovalenko:2003-CPL, Taminura:2007, Kovalenko:2004, Kovalenko:2017}. It enabled the modelling of sorption and supercapacitance of electrolyte solutions in functionalized nanoporous carbon electrodes. That included solvent-specific wetting and water depletion in hydrophobic carbon nanopores, asymmetry in solvation and adsorption of cations and anions, specific adsorption in functionalized carbon nanopores, desalination of ions in hydrophobic nanopores, and removal of ionic impurities from an aqueous waste stream in a nanoporous electrosorption cell.

The replica formalism describes ``annealed'' fluid of species 1 at equilibrium temperature $T_1$ sorbed in a ``quenched'' nanoporous matrix of species 0 described with an equilibrium ensemble at temperature $T_0$. The average free energy of the annealed fluid sorbed in the matrix is given by a statistical average of the free energy of the fluid over all distributions of matrix configurations ${\rm\bf q}_0$,
\begin{equation}
  \overline{A}_1 = A_1^{\rm id} - k_{\rm B}T_1 \bigl\langle \ln Z_1\left({\rm\bf q}_0\right) \bigr\rangle_{{\rm\bf q}_0} ,
  \label{eqn:A-ann-quench-gen}
\end{equation}
where $A_1^{\rm id}$ is the ideal gas free energy and $Z_1\left({\rm\bf q}_0\right)$ is the canonical partition function. The logarithm statistical average follows from the replica identity similar to theory of spin glasses relating it to the analytic continuation of moments $Z^s$: $\ln Z_1 = \lim_{s\to 0} {\rm d}Z^s/{\rm d}s$. The statistical average of the moments is given by the equilibrium canonical partition function of a fully annealed $(s+1)$-component liquid mixture of matrix species 0 and $s$ equivalent replicas of fluid species 1 not interacting with each other. The average free energy of the annealed fluid is obtained assuming no replica symmetry breaking in the analytic continuation of the annealed replicated free energy $A_{\rm rep}(s)$,
\begin{equation}
  \overline{A}_1 = \lim_{s\to 0} \frac{{\rm d}A_{\rm rep}(s)}{{\rm d}s} .
  \label{eqn:A-ann-quench-rep-lim}
\end{equation}

The replica Ornstein-Zernike integral equations for a quenched-annealed atomic system \cite{Given:1992-1,Given:1992-2, Given:1994} were extended to the replica DRISM integral equations for annealed associating molecular liquid sorbed in a quenched matrix \cite{Kovalenko:2001-1, Kovalenko:2001-2, Kovalenko:2003-CPL, Taminura:2007, Kovalenko:2004, Kovalenko:2017},
\begin{subequations} \label{eqn:replica-drism}
  \begin{align}
    h_{\alpha\gamma}^{00}(r) & = \omega_{\alpha\mu}^{00}(r)*c_{\mu\nu}^{00}(r)*\omega_{\nu\gamma}^{00}(r)
                                 + \ \omega_{\alpha\mu}^{00}(r)*c_{\mu\nu}^{00}(r)*\rho_\nu^0 h_{\nu\gamma}^{00}(r) , \label{eqn:replica-drism-A} \\
    h_{\alpha\gamma}^{10}(r) & = \tilde{\omega}_{\alpha\mu}^{11}(r)*c_{\mu\nu}^{10}(r)*\omega_{\nu\gamma}^{00}(r) 
                                 + \ \tilde{\omega}_{\alpha\mu}^{11}(r)*c_{\mu\nu}^{10}(r)*\rho_\nu^0 h_{\nu\gamma}^{00}(r) \notag \\
                             &   + \ \tilde{\omega}_{\alpha\mu}^{11}(r)*c_{\mu\nu}^{(\rm c)}(r)*\rho_\nu^1 h_{\nu\gamma}^{10}(r) , \label{eqn:replica-drism-B} \\
    h_{\alpha\gamma}^{01}(r) & = \omega_{\alpha\mu}^{00}(r)*c_{\mu\nu}^{01}(r)*\tilde{\omega}_{\nu\gamma}^{11}(r) 
                                 + \ \omega_{\alpha\mu}^{00}(r)*c_{\mu\nu}^{00}(r)*\rho_\nu^0 h_{\nu\gamma}^{01}(r) \notag \\
                             &   + \ \omega_{\alpha\mu}^{00}(r)*c_{\mu\nu}^{01}(r)*\rho_\nu^1 \tilde{h}_{\nu\gamma}^{(\rm c)}(r) , \label{eqn:replica-drism-C} \\
    \tilde{h}_{\alpha\gamma}^{11}(r) & = \tilde{\omega}_{\alpha\mu}^{11}(r)*c_{\mu\nu}^{11}(r)*\tilde{\omega}_{\nu\gamma}^{11}(r) 
                                         + \ \tilde{\omega}_{\alpha\mu}^{11}(r)*c_{\mu\nu}^{10}(r)*\rho_\nu^0 h_{\nu\gamma}^{01}(r) \notag \\
                                     &   + \ \tilde{\omega}_{\alpha\mu}^{11}(r)*c_{\mu\nu}^{(\rm c)}(r)*\rho_\nu^1 \tilde{h}_{\nu\gamma}^{10}(r)  
                                         + \ \tilde{\omega}_{\alpha\mu}^{11}(r)*c_{\mu\nu}^{(\rm b)}(r)*\rho_\nu^1 \tilde{h}_{\nu\gamma}^{(\rm c)}(r) , \label{eqn:replica-drism-D} \\
    \tilde{h}_{\alpha\gamma}^{(\rm c)}(r) & = \tilde{\omega}_{\alpha\mu}^{11}(r)*c_{\mu\nu}^{(\rm c)}(r)*\tilde{\omega}_{\nu\gamma}^{11}(r) 
                                              + \ \tilde{\omega}_{\alpha\mu}^{11}(r)*c_{\mu\nu}^{(\rm c)}(r)*\rho_\nu^1 \tilde{h}_{\nu\gamma}^{(\rm c)}(r) , \label{eqn:replica-drism-E} 
  \end{align}
\end{subequations}
where $\rho_\gamma^1$ and $\rho_\gamma^0$ are the site number densities of liquid species and matrix nanoparticles, $h_{\alpha\gamma}^{ij}(r)$ and $c_{\alpha\gamma}^{ij}(r)$ are the replica total and direct correlation functions between interaction sites $\alpha$ and $\gamma$ of species $i,j=1$ for liquid molecules and 0 for matrix nanoparticles). The replica liquid-liquid total and direct correlation functions have the connected (c) and disconnected, or blocking (b), portions,
\begin{subequations} \label{eqn:replica-RISM-blocking}
  \begin{align}
    \tilde{h}_{\alpha\gamma}^{11}(r) = \tilde{h}_{\alpha\gamma}^{(\rm c)}(r) + h_{\alpha\gamma}^{(\rm b)}(r) , \\ 
    c_{\alpha\gamma}^{11}(r) = c_{\alpha\gamma}^{(\rm c)}(r) + c_{\alpha\gamma}^{(\rm b)}(r) . 
  \end{align}
\end{subequations}
The connected correlations marked with superscript follow from the correlations between the same replica of the liquid, and the blocking ones from those between different replicas of the liquid in the analytical continuation limit $s\to\infty$. The blocking correlations are a subset of liquid-liquid Mayer diagrams with all paths between two root vortices passing through at least one field vortex of the matrix that are completely blocked by matrix vortices (indirect, matrix-mediated part of the liquid-liquid correlations). The remaining part is the connected liquid-liquid correlations.

A closure relation between the total and direct correlation functions complements the replica RISM integral equations~(\ref{eqn:replica-KH-closure}) to provide a computational handle to integrate the infinite chain of diagrams. Though the exact closure can be formally expressed as a series in terms of multiple integrals of the combinations of the total correlation functions, it is cumbersome, and in practice is replaced with tenable approximations. Among a number of closure relations, the Kovalenko-Hirata (KH) closure was shown to work for small molecules, biomolecules, nanomaterials, aggregates, ligand-protein interactions, with high accuracy in the whole range of thermodynamic conditions \cite{Johnson:2016, Kondratenko:2015, Genheden:2014, Stoyanov:2013, Stoyanov:2008, Imai:2005}. It consistently accounts for both electrostatic, associative, and steric effects of solvation in simple and complex liquids.

The KH closure applied to the matrix-matrix, liquid-matrix, and liquid-liquid correlations has the form
\begin{equation}
  \begin{split}
    g_{\alpha\gamma}^{ij}(r) = & \begin{cases}
      \exp\left( d_{\alpha\gamma}^{ij}(r) \right) & \text{for } d_{\alpha\gamma}(r) \leqslant 0 \\
	  1 + d_{\alpha\gamma}^{ij}(r) / & \text{for } d_{\alpha\gamma}(r) > 0 
    \end{cases} , 
    \\
    d_{\alpha\gamma}^{ij}(r) = & - u_{\alpha\gamma}^{ij}(r) / \left(k_{\rm B}T\right) + h_{\alpha\gamma}^{ij}(r) - c_{\alpha\gamma}^{ij}(r) , 
  \end{split}
  \label{eqn:replica-KH-closure}
\end{equation}
where $u_{\alpha\gamma}^{ij}(r)$ is the interaction potential between interaction sites $\alpha$ and $\gamma$ of species $i$ and $j$ specified by a molecular force field. The KH closure (\ref{eqn:replica-KH-closure}) couples in a nontrivial way the mean spherical approximation (MSA) applied to spatial regions of solvent density enrichment $g>1$ and the hypernetted chain (HNC) closure applied to the regions of density depletion $g<1$ approximation. The solvent site distribution function and its first derivative are continuous at the joint boundary $g=1$ by construct. The RISM-HNC theory is known to overestimate the solvation structures in strongly associated system thus imparting numerical instability and also diverges for strongly charged systems such as electrolytes in a solution. The RISM-KH theory can easily handle a strongly associated system. Other closure relations developed over the years are aimed at specific applications, lacking the generality of the KH closure \cite{Percus:1958, Martynov:1983, Ballone:1986, Kast:2008}. Though the KH closure underestimates the height of strong associative peaks, it widens the peak to some extent and so provides a correct thermodynamics and solvation structure \cite{Perkyns:2011, Stumpe:2011}.

The modified Verlet (VM) closure is applied to the blocking part of the correlations,
\begin{subequations} \label{eqn:replica-VM-closure}
  \begin{align}
    g_{\alpha\gamma}^{(\rm b)}(r) = & \ h_{\alpha\gamma}^{(\rm b)}(r) + 1 = \exp\left( h_{\alpha\gamma}^{(\rm b)}(r) - c_{\alpha\gamma}^{(\rm b)}(r) + b_{\alpha\gamma}^{(\rm b)}(r) \right) , \label{eqn:replica-VM-closure-A} \\
    b_{\alpha\gamma}^{(\rm b)}(r) = & - \frac{1}{2} \frac{\left( t_{\alpha\gamma}^{(\rm b)}(r) \right)^2} {1+a\max\left( t_{\alpha\gamma}^{(\rm b)}(r), 0 \right)} . \label{eqn:replica-VM-closure-B}
  \end{align}
\end{subequations}
The VM bridge correction (\ref{eqn:replica-VM-closure-B}) was expressed in terms of the nodal correlation function, with the parameter $a=0.8$ the same as in the original Verlet correction. The blocking correlations closure (\ref{eqn:replica-VM-closure}) does not have an interaction potential, since different replicas of fluid do not interact with each other in the limit $s\to\infty$. This blocking approximation (\ref{eqn:replica-VM-closure}) appropriately represents nonlinear blocking correlations in polar solvents as well as electrolyte solutions in nanoporous materials, either neutral or charged. With the analytical treatment of the electrostatic asymptotics similarly to the bulk DRISM-KH equations (\ref{eqn:drism})--(\ref{eqn:chi}), the replica DRISM-VM integral equations are converged using the MDIIS accelerated numerical solver~\cite{Kovalenko:1999-MDIIS}.

The excess chemical potential of the liquid sorbed in disordered nanopores was decomposed into the host matrix contributions from to the liquid-matrix $(i,j=1,0)$ correlations and the sorbed liquid part from the liquid-liquid $(i,j=1,1)$ correlations minus an additional term from the blocking (b) correlations~\cite{Kovalenko:2001-1, Kovalenko:2001-2},
\begin{equation}
  \Delta\mu_s^{1} = \Delta\mu_s^{10} + \Delta\mu_s^{11} - \Delta\mu_s^{\rm (b)} .
  \label{eqn:replica-RISM-ex-chem-pot}
\end{equation}
Like equations (\ref{eqn:SFE})--(\ref{eqn:chi}), the replica DRISM integral equations (\ref{eqn:replica-drism}) with the replica KH closure (\ref{eqn:replica-KH-closure}) and replica VM closure (\ref{eqn:replica-VM-closure}) gives the terms of the excess chemical potential in a closed analytical form of the liquid-matrix and liquid-liquid correlation functions for liquid ($j=1$) and matrix ($j=0$) species. The blocking correlations term in the chemical potential is obtained from the relations between the connected and blocking parts replica DRISM equations (\ref{eqn:replica-RISM-blocking}) with the replica VM closure (\ref{eqn:replica-VM-closure}) using thermodynamic integration as
\begin{eqnarray}
  \Delta\mu_s^{\rm (b)} &=& k_{\rm B}T \sum_{\alpha,\gamma\in 1} \rho_\alpha^1 4\piup \int_0^\infty {\rm d}r\, r^2  \left[ \frac{1}{2} \left( h_{\alpha\gamma}^{\rm (b)}(r) \right)^2 - \frac{1}{2} h_{\alpha\gamma}^{\rm (b)}(r) c_{\alpha\gamma}^{\rm (b)}(r) \nonumber\right.\\
  &+&\left. \left( h_{\alpha\gamma}^{\rm (b)}(r) + 1 \right) b_{\alpha\gamma}^{\rm (b)}(r) - s_{\alpha\gamma}^{\rm (b)}(r) \right] ,
  \label{eqn:replica-RISM-ex-chem-pot-blocking}
\end{eqnarray}
with the star function derived assuming a unique functionality of the correlations \cite{Kovalenko:2001-1}
\begin{equation}
  s_{\alpha\gamma}^{\rm (b)}(r) = \begin{cases}
  	- \dfrac{1}{2a^3} \left[ \ln\left(1+a t_{\alpha\gamma}^{\rm (b)}(r)\right) - a t_{\alpha\gamma}^{\rm (b)}(r) + \dfrac{1}{2}\left(a t_{\alpha\gamma}^{\rm (b)}(r) \right)^2 \right] 
  	  \dfrac{t_{\alpha\gamma}^{\rm (b)}(r)}{h_{\alpha\gamma}^{\rm (b)}(r)}  & \text{for } \ t_{\alpha\gamma}^{\rm (b)}(r) > 0 \\
  	- \dfrac{1}{6} t_{\alpha\gamma}^{\rm (b)}(r) h_{\alpha\gamma}^{\rm (b)}(r)  & \text{for } \ t_{\alpha\gamma}^{\rm (b)}(r) \leqslant 0 
  \end{cases} .
  \label{eqn:replica-RISM-ex-chem-pot-blocking-star-function}
\end{equation}
Compressibility and other thermodynamic derivatives of the sorbed annealed solution have an exact analytical form of the connected terms of correlations.

In the electrolyte solution sorbed at the surface of nanopores, an EDL forms even in the absence of an external electric charge due to the orientation of polar solvent molecules at the surface and the asymmetry of the cationic and anionic density distributions. For interaction site charges $q_f^0$ of chemical functional groups of sort $f\in 0$ grafted on the surface of a matrix nanoparticle of sort $c\in 0$, the statistical charge density distribution $\tau_c^0$ of interaction site charges $q_\gamma^1$ around the labeled functionalized matrix nanoparticle is as follows:
\begin{equation}
  \tau_c^0(r) = \sum_{f\in 0} q_f^0 \rho_f^0 g_{cf}^{00}(r) + \sum_{\gamma\in 1} q_\gamma^1 \rho_\gamma^1 g_{c\gamma}^{01}(r) ,   
  \label{eqn:replica-RISM-charge-density}
\end{equation}
where both matrix nanoparticles $c$ and chemical functional groups $f$ grafted to them belong to the matrix species $j=0$. 

The statistical charge density distribution $\tau_c^0$ corresponds to the statistically averaged full local electrostatic potential $\mathit{\Psi}_c^0(r)$ around a matrix nanoparticle $c$ according to the Poisson equation
\begin{equation}
  \Delta^2 \psi_c^0(r) = -4\piup\tau_c^0(r) . 
  \label{eqn:Poisson}
\end{equation}
The full electrostatic potential $\phi_c^0(r)$ also includes the term from an externally induced charge $q_c^0$ on the conducting matrix nanoparticle of radius $R_c^0$,
\begin{equation}
  \phi_c^0(r) = \psi_c^0(r) + \frac{q_c^0}{\max\left(r,R_c^0\right)} .
  \label{eqn:full-electrostatic-potential}
\end{equation}
The external charge density
\begin{equation}
  q_{\rm ext} = \sum_{c\in 0} q_c^0 \rho_c^0 
  \label{eqn:ext-charge}
\end{equation}
on the electrode and the opposite external charge $-q_{\rm ext}$ on the other electrode of the supercapacitor result in separation of electrolyte cations and anions that diffuse across the electrodes separator to the electric double layers in their nanopores. This diffusion occurs until the ionic concentrations bias in each electrode satisfies the condition of electroneutrality in the whole system,
\begin{equation}
  q_{\rm ext} + \sum_f q_f^0 \rho_f^0 + \sum_\gamma q_\gamma^1 \rho_\gamma^1 = 0 . 
  \label{eqn:electroneutrality}
\end{equation}
The diffusion exchange between the electrodes and the bulk solution bath adjust the bias between the densities of cations and anions until they satisfy the electroneutrality in each electrode. For connected carbon nanospheres of sorts $\alpha$ and sizes $R_c^0$, the external charge $q_{\rm ext2}$ is distributed among the charges $q_c^0$ on each sort of nanospheres, provided the electrostatic potential is the same inside all carbon nanospheres:
\begin{equation}
  \phi_c(r) \equiv \phi_c^0(r<R_c^0;q_{\rm ext}) \quad {\rm for} \ {\rm all} \quad c\in 0 .  
  \label{eqn:electrostati-potential}
\end{equation}
Further, the electrostatic potential far from each nanosphere of sort $c$ gives the average electrode potential with both the carbon nanospheres and sorbed solution parts,
\begin{equation}
	\phi_{\rm av}(r) \equiv \phi_c^0(r=\infty_c^0;q_{\rm ext}) \quad {\rm for} \ {\rm all} \quad c\in 0 .  
	\label{eqn:electrostati-potential-infty}
\end{equation}
The electrostatic potential of the carbon electrode is given by the potential of each carbon nanosphere $\phi_c(q_{\rm ext})$. The potential change from the carbon nanoparticle to the solution outside the electrode consists of the following parts: (i) Voltage across the electric field from $\phi_c(q_{\rm ext})$ at the intrinsic surface of nanopores to the average level $\phi_{\rm av}(q_{\rm ext})$ of all other carbon nanoparticles in the nanoporous material, and (ii) Via the electric field $\phi_{\rm ext-EDL}$ of the external EDL at the macroscopic surface of the electrode to the solution bulk outside the electrode. The ``zero'' potential level is therefore shifted from the potential in vacuum by the external EDL $\phi_{\rm ext-EDL}$. 

For sorbed species $s$ with density $\rho_s$, the chemical potential comprises (1) the ideal gas term of spheres with molecular weight $m_s$, 
$\Delta m_s^{\rm id} = k_{\rm B}T \ln\left( \rho_s \Lambda_s \right)$, where $\Lambda_s = \sqrt{h^2/\left( 2\piup m_s k_{\rm B}T \right)}$ 
is the de Broglie thermal wave length, the excess chemical potential $\Delta\mu_s$ for liquid-liquid and liquid-matrix interactions in the nanopores, and (2) the electrostatic potential of the species charges $q_s$ in the average electrostatic field between the electrodes, 
\begin{equation}
  \mu_s = \mu_s^{\rm id} + \Delta\mu_s + q_s \phi_{\rm av} .
  \label{eqn:chemical-potential}
\end{equation}
These chemical potential terms arise, respectively, from the osmotic effect, the interaction in the nanopores, and the ``zero'' level of the nanoporous electrode relative to the vacuum potential following from the Nernst equation [74].

Inside a charged electrode, the excess chemical potentials of ionic species strongly differ from the bulk solution, which causes diffusion of ions across the separator to have the electric field of ionic dipoles at the electrodes boundaries counterbalance the difference of the “interior” chemical potential $\mu_s^{\rm id} + \Delta\mu_s$ of electrodes I and II, and to equalize the chemical potential in the electrodes, $\mu_s^{\rm I} = \mu_s^{\rm II}$. The bias between the statistically averaged ``zero'' levels of the two electrodes $\phi_{\rm av}^{\rm II}(q_{\rm ext})$ and $\phi_{\rm av}^{\rm I}(-q_{\rm ext})$  with opposite external charges $\pm q_{\rm ext}$ is obtained as
\begin{equation}
  q_s \left( \phi_{\rm av}^{\rm II}(q_{\rm ext}) - \phi_{\rm av}^{\rm I}(-q_{\rm ext}) \right) = k_{\rm B}T \ln\left( \rho_s^{\rm I} / \rho_s^{\rm II} \right) + \Delta\mu_s^{\rm I}(-q_{\rm ext}) - \Delta\mu_s^{\rm II}(q_{\rm ext}) .
  \label{eqn:chemical-equilibrium-bias}
\end{equation}
The same bias of the electrostatic potential levels $\phi_{\rm av}^{\rm II} - \phi_{\rm av}^{\rm I}$ must satisfy equation (\ref{eqn:chemical-equilibrium-bias}) for all the solution species. Diffusion of the ionic species to the corresponding electrode with the lower chemical potential occurs due to a difference between the bias values necessary to counterbalance the ``interior'' chemical potential $\mu_s^{\rm id} + \Delta\mu_s$ until the chemical equilibrium (\ref{eqn:chemical-equilibrium-bias}) is reached for both cations and anions. Moreover, neutral solvent molecules diffuse between the two electrodes until their osmotic term and excess chemical potentials satisfy the bias condition (\ref{eqn:chemical-equilibrium-bias}), which reduces to the equality of their ``interior'' parts $\mu_s^{\rm id} + \Delta\mu_s$.

The supercapacitor device voltage obtained by summing up the statistically averaged electrostatic potential changes: (i) across the device consists of the potential changes in the electrode I intrinsic EDL, $\phi_{\rm av}^{\rm I}(-q_{\rm ext}) - \phi_c^{\rm I}(-q_{\rm ext})$; (i) from the ``zero'' level of electrode I to the solution bulk and then to the ``zero'' level of electrode II, $\phi_{\rm av}^{\rm II}(q_{\rm ext}) - \phi_{\rm av}^{\rm I}(-q_{\rm ext})$; and (iii) in the electrode II intrinsic EDL, $\phi_c^{\rm II}(q_{\rm ext}) - \phi_{\rm av}^{\rm II}(q_{\rm ext})$. With the relation~(\ref{eqn:average-electrostatic-potential}) for the average electrostatic potentials in terms of the chemical potentials and number densities of sorbed ions, the supercapacitor voltage assumes the form
\begin{equation}
  U(q_{\rm ext}) = \left( \phi_{\rm av}^{\rm I} - \phi_c^{\rm I} \right) - \left( \phi_{\rm av}^{\rm II} - \phi_c^{\rm II} \right) - \frac{1}{q_s} \left( k_{\rm B}T \ln\left( \rho_s^{\rm I} / \rho_s^{\rm I}/\right) + \Delta\mu_s^{\rm I} - \Delta\mu_s^{\rm II} \right) .
  \label{eqn:average-electrostatic-potential}
\end{equation}
The chemical equilibrium conditions (\ref{eqn:chemical-equilibrium}) should be converged for fluid densities $\rho_s$, with the replica DRISM-KH-VM integral equations (\ref{eqn:replica-drism})--(\ref{eqn:replica-VM-closure}) being converged at each outer loop of iterations for $\rho_s$. This is followed by calculation of the potential changes $\phi_{\rm av}^{\rm I} - \phi_c^{\rm I}$ and $\phi_{\rm av}^{\rm II} - \phi_c^{\rm II}$ at converged $\rho_s$ by solving the Poisson equation (\ref{eqn:Poisson}).

The purification efficiency relation for an electrosorption cell which holds sorbed electrolyte with ionic concentrations inside electrodes I and II at a much lower bulk electrolyte concentration $\rho_s^{\rm blk}$, and an excess chemical potential $\Delta\mu_s^{\rm blk}$ in the bulk solution efflux is obtained from the chemical equilibrium condition
\begin{subequations} \label{eqn:chemical-equilibrium}
  \begin{align}
    \frac{\rho_s^{\rm blk}}{\rho_s^{\rm I}} = & \exp\left[ -\frac{1}{k_{\text{B}}T} \left(\Delta\mu_s^{\rm blk} - \Delta\mu_s^{\rm I}(-q_{\rm ext}) - \Delta\mu_{\rm av}^{\rm I}(-q_{\rm ext})\right) \right] \quad {\rm for} \; q_s > 0 \; {\rm (cations)} , \\
    \frac{\rho_s^{\rm blk}}{\rho_s^{\rm II}} = & \exp\left[ -\frac{1}{k_{\text{B}}T} \left(\Delta\mu_s^{\rm blk} - \Delta\mu_s^{\rm II}(-q_{\rm ext}) - \Delta\mu_{\rm av}^{\rm II}(-q_{\rm ext})\right) \right] \quad {\rm for} \; q_s < 0 \; {\rm (anions)} .
  \end{align}
\end{subequations}
Select applications of the developments in the replica RISM-KH-VM molecular solvation theory are briefly described below.

\section{Aqueous electrolyte solution of lithium, potassium, and manganese hydroxides in a nanoporous carbon supercapacitor}

\setcounter{equation}{38}

\begin{figure}[!t]
	\centering
	\includegraphics[width=0.6\textwidth]{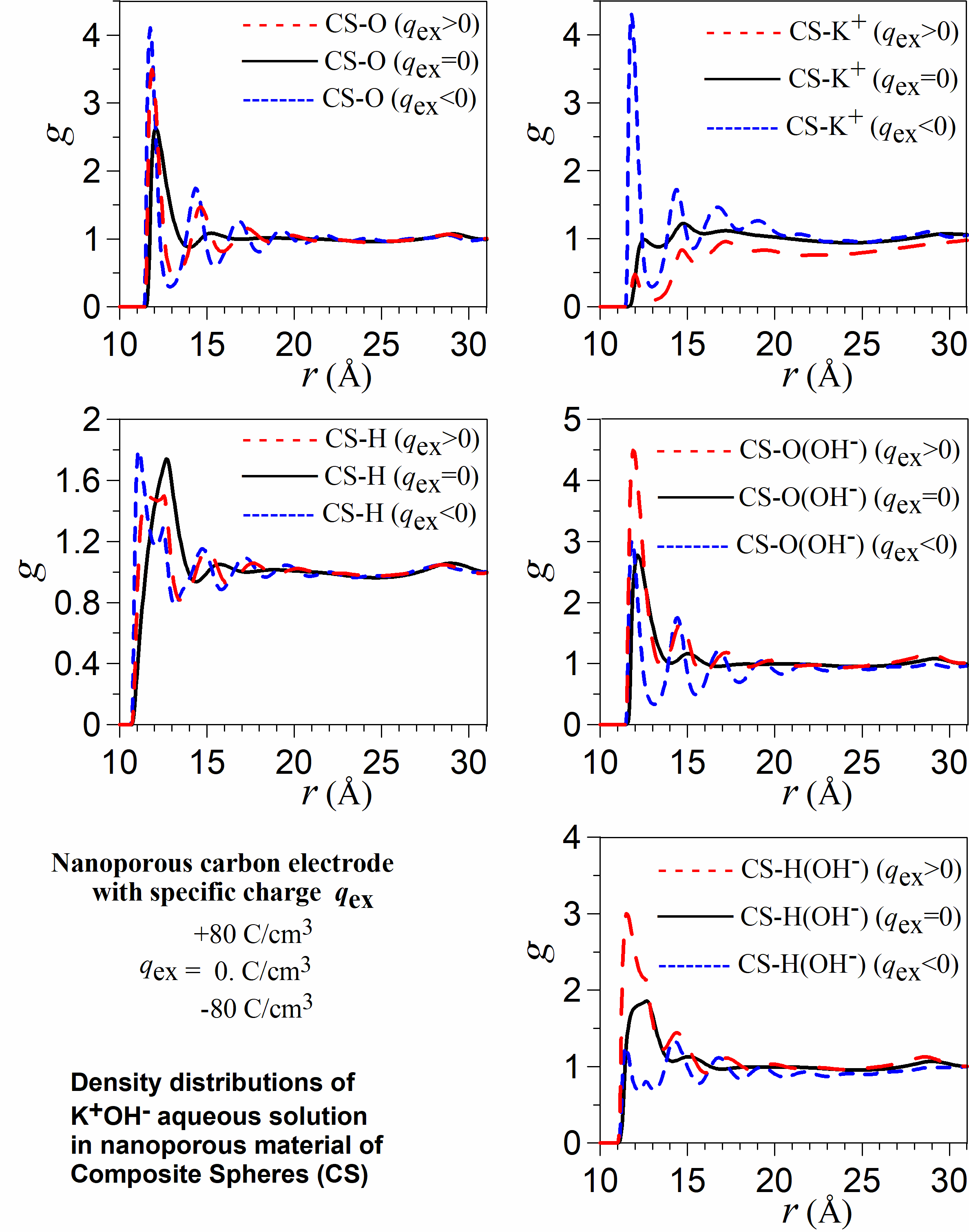}
	\caption{(Colour online) Solvation structure of KOH aqueous electrolyte solution sorbed in the nanoporous carbon electrode. RDFs of water O and H sites, and of K$^+$ and OH$^-$ ions around carbon nanoparticles. Nanoporous electrode charges: $q_{\rm ext}=0$ (solid black lines); $q_{\rm ext}=+80$ C/cm$^3$ (long-dashed red lines); $q_{\rm ext}=-80$ C/cm$^3$ (short-dashed blue lines).}
	\label{fig:CS-KOH-g(r)}
\end{figure}

The replica RISM-KH-VM modelling of KOH aqueous electrolyte solution in a nanoporous carbon supercapacitor revealed that its solvation structure and thermodynamics are profoundly different from planar electrode-electrolyte interfaces. Figure \ref{fig:CS-KOH-g(r)} presents the changes in the carbon-solution radial distribution functions (RDFs) of KOH aqueous electrolyte solution at ambient thermodynamics conditions in the nanoporous carbon electrode with charge varying from 0 to $q_{\rm ext} = \pm 80$ C/cm$^3$. The external charge strongly affects the RDFs between carbon nanospheres and ions, attracting counterions and repelling co-ions. The carbon-water oxygen and hydrogen distributions strongly enhance with electrode charge, both positive and negative, exhibiting water attraction to carbon nanospheres with their charge. Moreover, water hydrogens strongly shift towards the nanopore surface for both positive and negative electrode charges. However, the RDFs between carbon nanoparticles and water oxygen and hydrogen barely change with the nanoporous carbon charge. 

\begin{figure}[!t]
	\centering
	\includegraphics[width=0.4\textwidth]{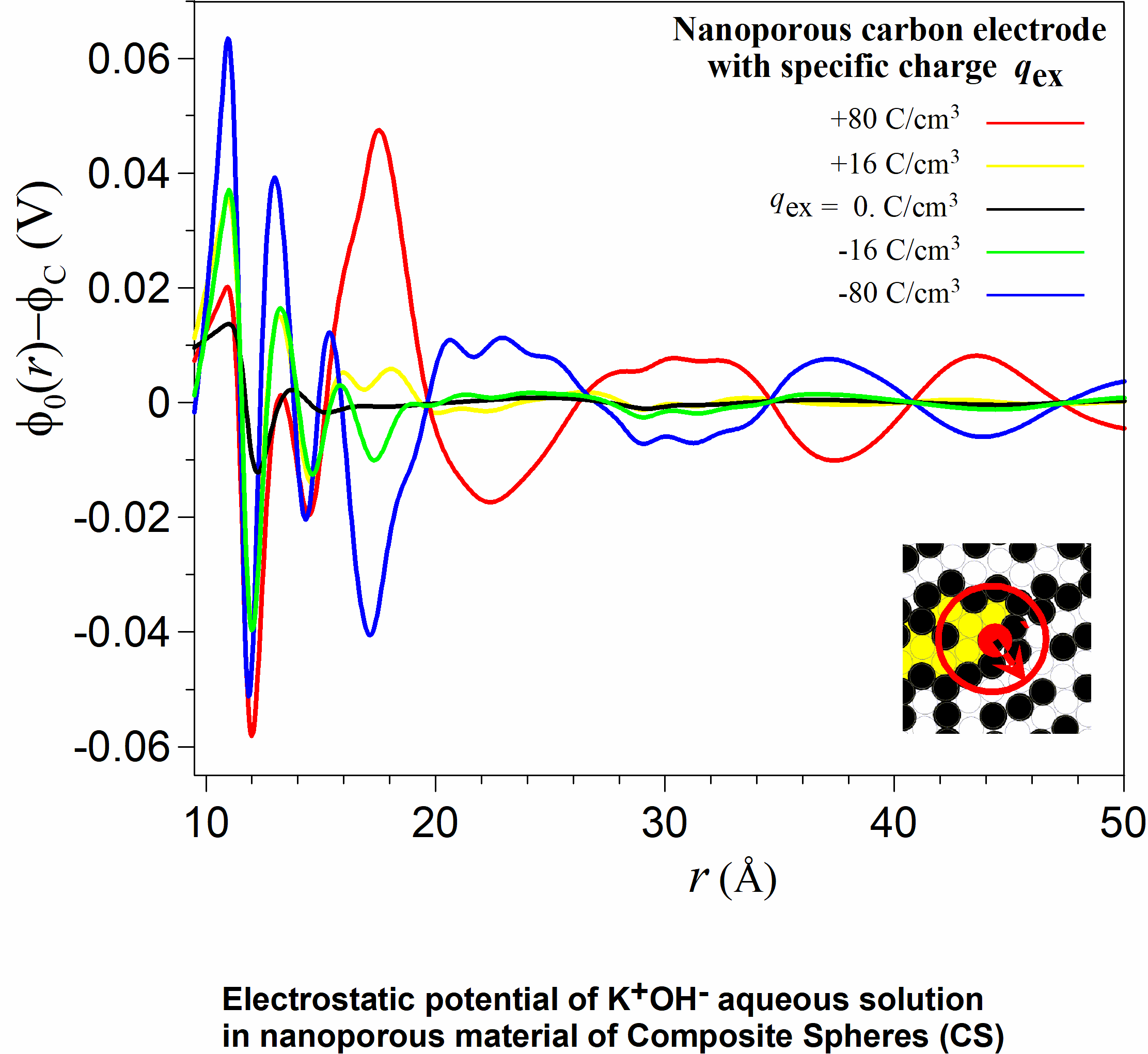}
	\caption{(Colour online) Electrostatic potential $\phi_0(r)$ around a nanoparticle of the nanoporous carbon electrode with respect to ``zero'' level $\phi_c$. The sorbed solution is in equilibrium with the bulk ambient aqueous solution of KOH electrolyte at concentration 120 ppm. Nanoporous electrode charges: $q_{\rm ext}=0$ C/cm$^3$ (black line); $q_{\rm ext}=+16$ C/cm$^3$ (yellow line); $q_{\rm ext}=+80$ C/cm$^3$ (red line); $q_{\rm ext}=-16$ C/cm$^3$ (green line); $q_{\rm ext}=-80$ C/cm$^3$ (blue line). Inset: Statistical mechanical average (red circle and distance vector) around a carbon nanoparticle (red ball) over carbon nanoparticles (black balls) and nanopores (white voids).}
	\label{fig:CS-KOH-fi(r)}
\end{figure}
\begin{figure}[!t]
	\centering
	\includegraphics[width=0.6\textwidth]{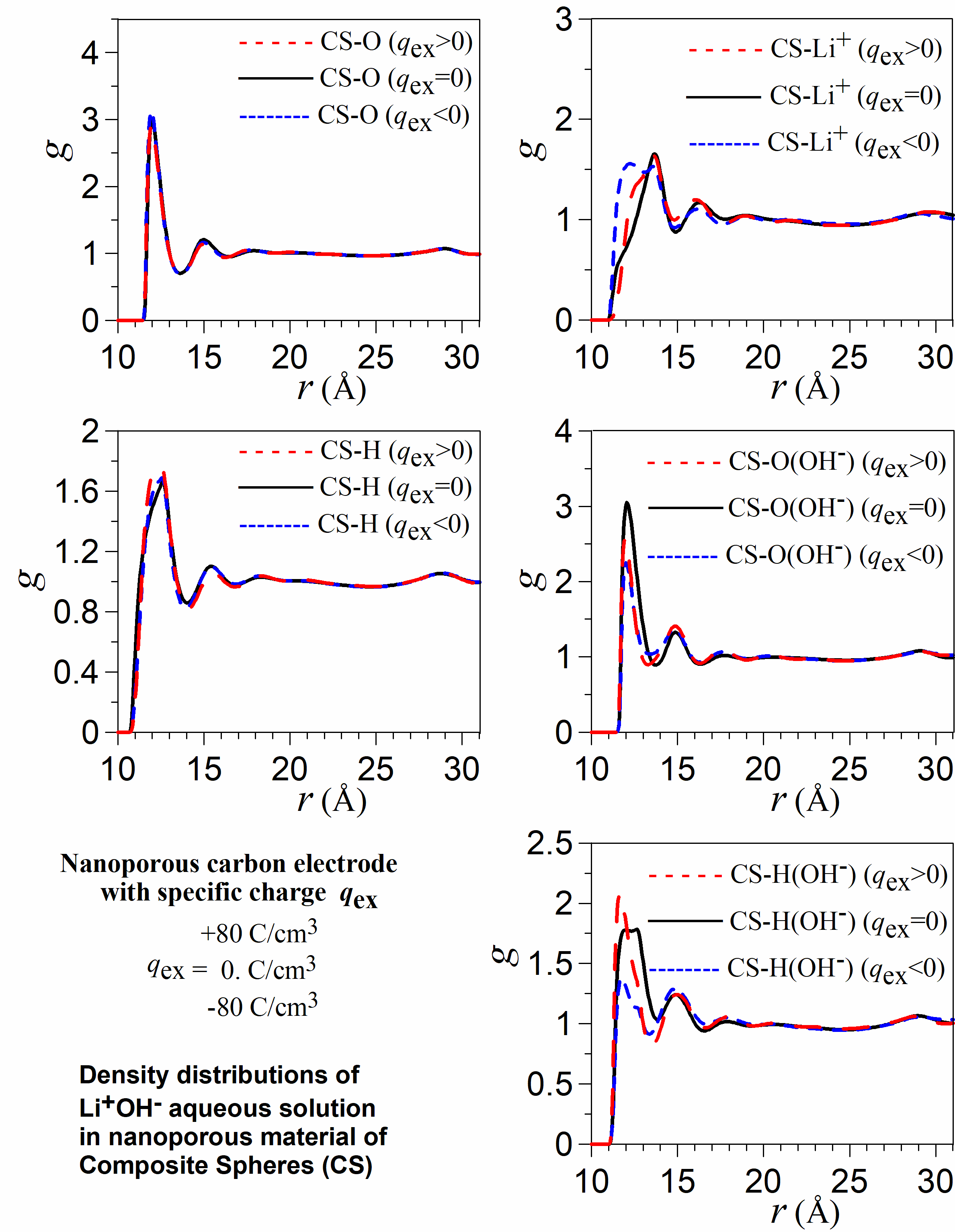}
	\caption{(Colour online) Solvation structure of LiOH aqueous electrolyte solution sorbed in the nanoporous carbon electrode. RDFs of water O and H sites, and of Li$^+$ and OH$^-$ ions around carbon nanoparticles. Nanoporous electrode charges are the same as in figure~\ref{fig:CS-KOH-g(r)}.}
	\label{fig:CS-LiOH-g(r)}
\end{figure}

\begin{figure}[!t]
	\centering
	\includegraphics[width=0.4\textwidth]{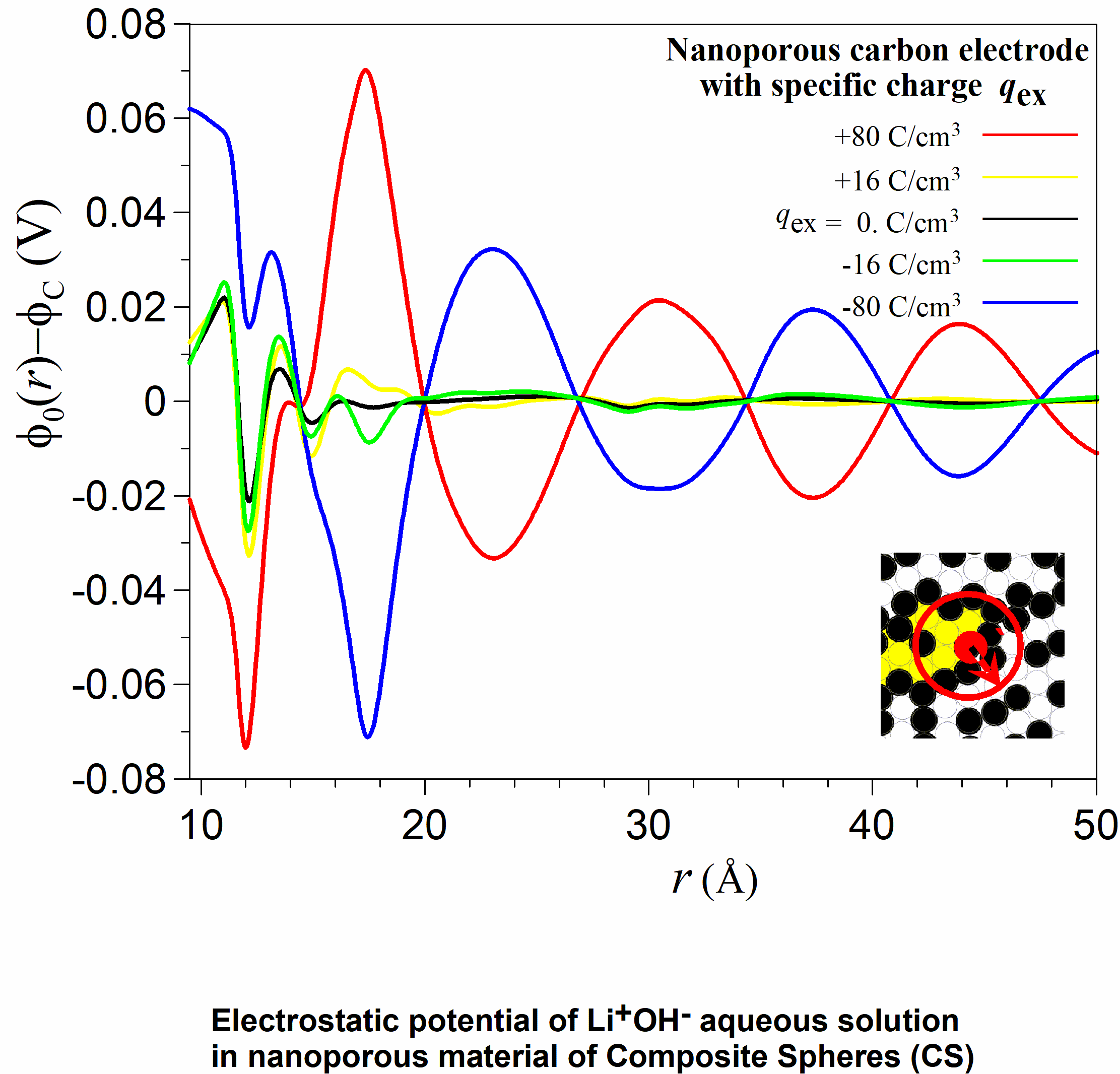}
	\caption{(Colour online) Electrostatic potential $\phi_0(r)$ around a nanoparticle of the nanoporous carbon electrode with respect to ``zero'' level $\phi_c$. The sorbed solution is in equilibrium with the bulk ambient aqueous solution of LiOH electrolyte at concentration 120 ppm. Nanoporous electrode charges and the Inset are the same as in figure \ref{fig:CS-KOH-fi(r)}.}
	\label{fig:CS-LiOH-fi(r)}
\end{figure}

Figure \ref{fig:CS-KOH-g(r)} depicts the run of the statistically averaged electrostatic potential $\psi_c^0(r)$ around a nanoparticle in the carbon nanoporous electrode obtained from the Poisson equation (\ref{eqn:Poisson}) for the electric charge densities $\tau_c^0$ with respect to the ``zero'' level $\phi_c$. The value of the electrostatic potential at $r<R_c^0$ inside the conducting carbon nanoparticle levels out. The electrostatic potential $\phi_0(r)$ varies from the surface of a carbon nanoparticle, through the intrinsic ELD at the nanoparticle surface, and to the bulk value $\phi_c$ inside the nanoporous electrode averaged over the nanoporous carbon material. The external charge on carbon nanoparticles steers the electrostatic potential in the first and second solvation shells near the nanoparticles surface. Solutions charges are absent in the Stern layer due to the steric constraints, and the slope of the potential curves for $r>R_c^0$  near the carbon nanoparticle surface is determined by the Coulomb potential. The surface dipole due to water hydrogens located closer to the nanoparticle surface than water oxygens causes the potential drop. Further, OH$^-$ ions located closer to the surface than K$^+$ ions cause the subsequent potential rise. The outer Helmholtz layer is formed by the peaks of the electrostatic potential in the first and second solvation layers. The electric charge of carbon nanoparticles makes the electrostatic potential oscillate with a period of 3 \AA \ close to the size of carbon nanoparticles, and extend to about 12 \AA \ from nanopores, slowly decaying with distance. This behavior includes both the nanoparticle diffuse layer and the EDL statistical average around all the closely packed nanoparticles of the nanoporous carbon. The potential change in the Stern layer is almost canceled out by the outer Helmholtz layer and further oscillations that come mostly from OH$^-$ ions in the two solvation shells screening the positive external charge and K$^+$ ions screening the negative external charge of the carbon nanoparticle.

The whole range of oscillations includes the diffuse layer around the nanoparticle, as well as the statistical average of the EDLs around other nanoparticles which are closely packed and correlated in the nanoporous carbon matrix. The potential drop of the Stern layer is almost completely cancelled out by the electric field of the outer Helmholtz layer and further oscillations which stems mainly from the ionic cloud of the first and second solvation shells screening the external charge of the carbon nanosphere, with OH$^-$ ions prevailing for positive, and K$^+$ for negative external charge. The model picks changes in the chemical potentials of both K$^+$ and OH$^-$ ions upon introduction of the EDL in the computation.
\begin{figure}[!t]
	\centering
	\includegraphics[width=0.6\textwidth]{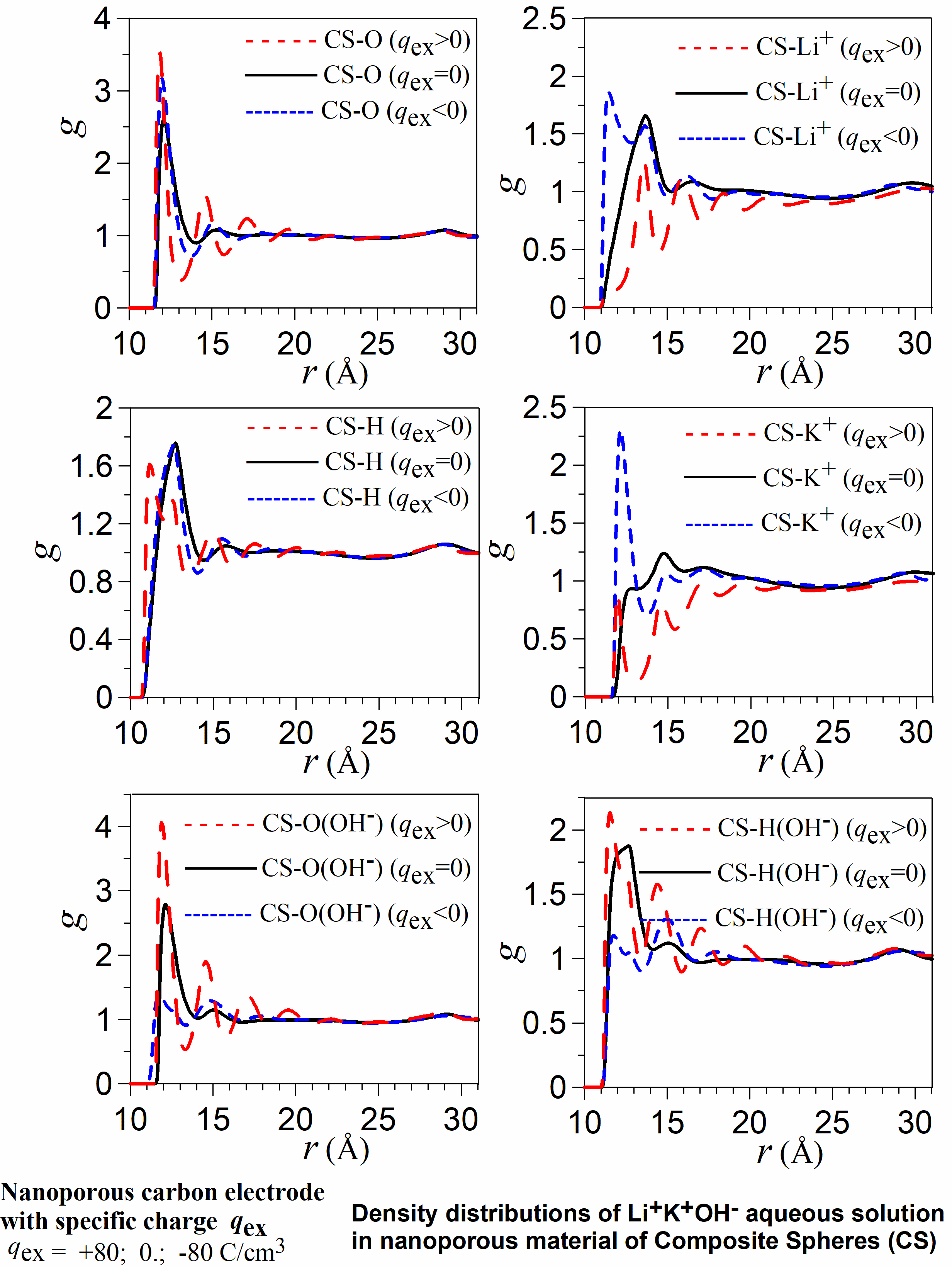}
	\caption{(Colour online) Solvation structure of aqueous solution of LiKOH electrolyte mixture sorbed in the nanoporous carbon electrode. RDFs of water O and H sites, and of Li$^+$, K$^+$ and OH$^-$ ions around carbon nanoparticles. Nanoporous electrode charges are the same as in figure \ref{fig:CS-KOH-g(r)}.}
	\label{fig:CS-LiKOH-g(r)}
\end{figure}

\begin{figure}[!t]
	\centering
	\includegraphics[width=0.43\textwidth]{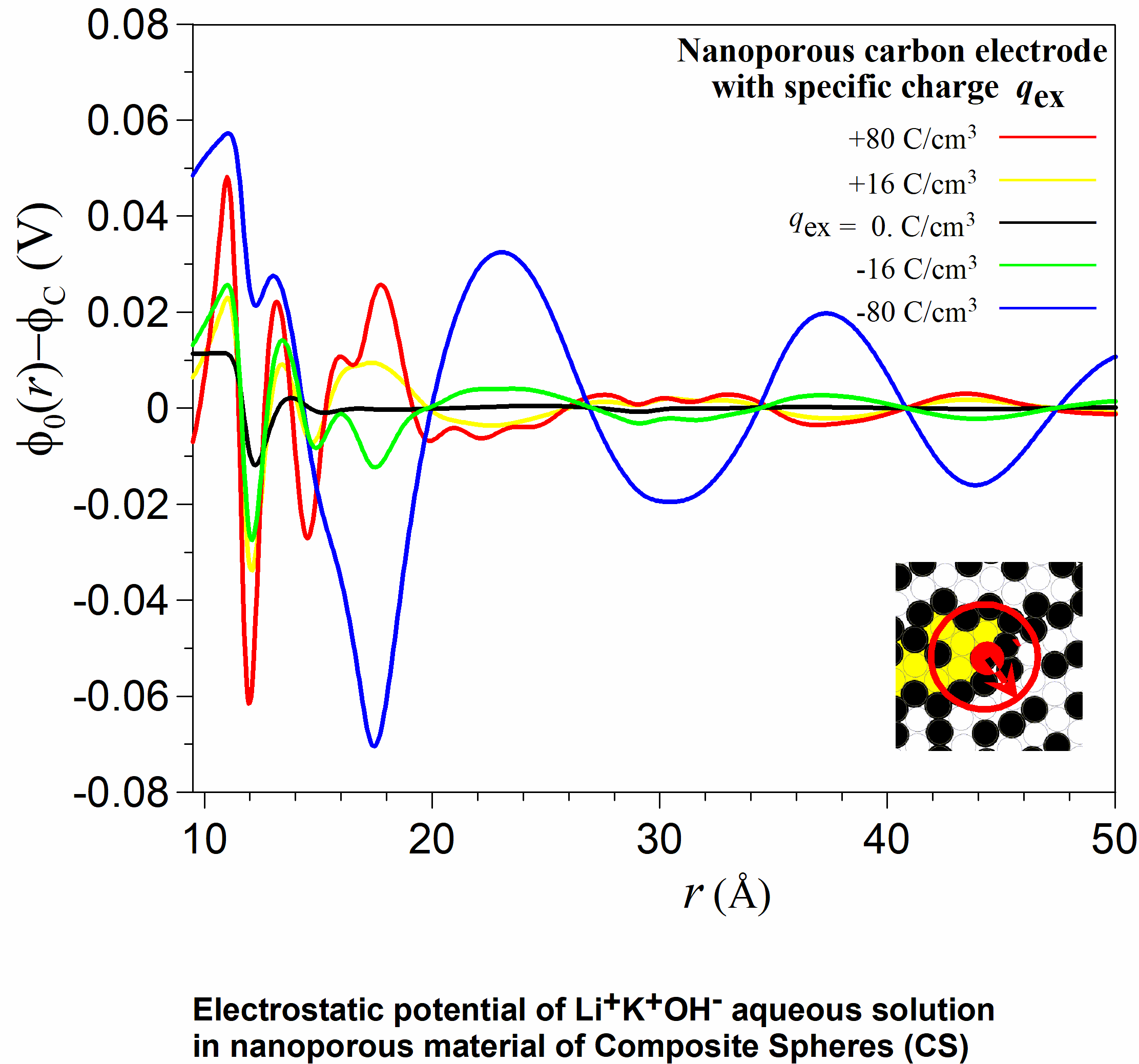}
	\caption{(Colour online) Electrostatic potential $\phi_0(r)$ around a nanoparticle of the nanoporous carbon electrode with respect to ``zero'' level $\phi_c$. The sorbed solution is in equilibrium with the bulk ambient aqueous solution of LiKOH electrolyte at concentration 120 ppm. Nanoporous electrode charges and the Inset are the same as in figure \ref{fig:CS-KOH-fi(r)}.}
	\label{fig:CS-LiKOH-fi(r)}
\end{figure}

For LiOH aqueous electrolyte solution sorbed in nanoporous carbon electrodes, the RDFs dramatically differ from those of KOH. Figure \ref{fig:CS-LiOH-g(r)} shows the carbon-solution RDFs of LiOH electrolyte at $q_{\rm ext}=0$ and $\pm 80$ C/cm$^3$. Unlike KOH electrolyte, the carbon-water RDFs for LiOH almost do not change with charging the nanopores, and the nanopore-ion RDFs significantly shift towards the nanopore surface, reflecting the attraction of both Li$^+$ and OH$^-$ ions with an increasing external charge, both positive and negative. The width of the RDF peak of Li$^+$ significantly increases for $q_{\rm ext}<0$ and weakly changes for $q_{\rm ext}>0$. The changes of the RDF peak of OH$^-$ ions with positive and negative external charge are opposite for H sites in OH$^-$ ions compared to Li$^+$; although quite similar for the O sites in OH$^-$ ions. The oscillations of all the RDFs between nanopores and Li$^+$ and OH$^-$ ions at the surface of nanopores are much less pronounced than those for K$^+$ and OH$^-$. Near the surface of nanopores, the oscillations of the electrostatic potential of Li$^+$ and OH$^-$ ions shown in figure \ref{fig:CS-LiOH-fi(r)} are much less pronounced than those for K$^+$ and OH$^-$, too, but are about twice larger at a distance from the surface of nanopores. 
\begin{figure}[!t]
	\centering
	\includegraphics[width=0.63\textwidth]{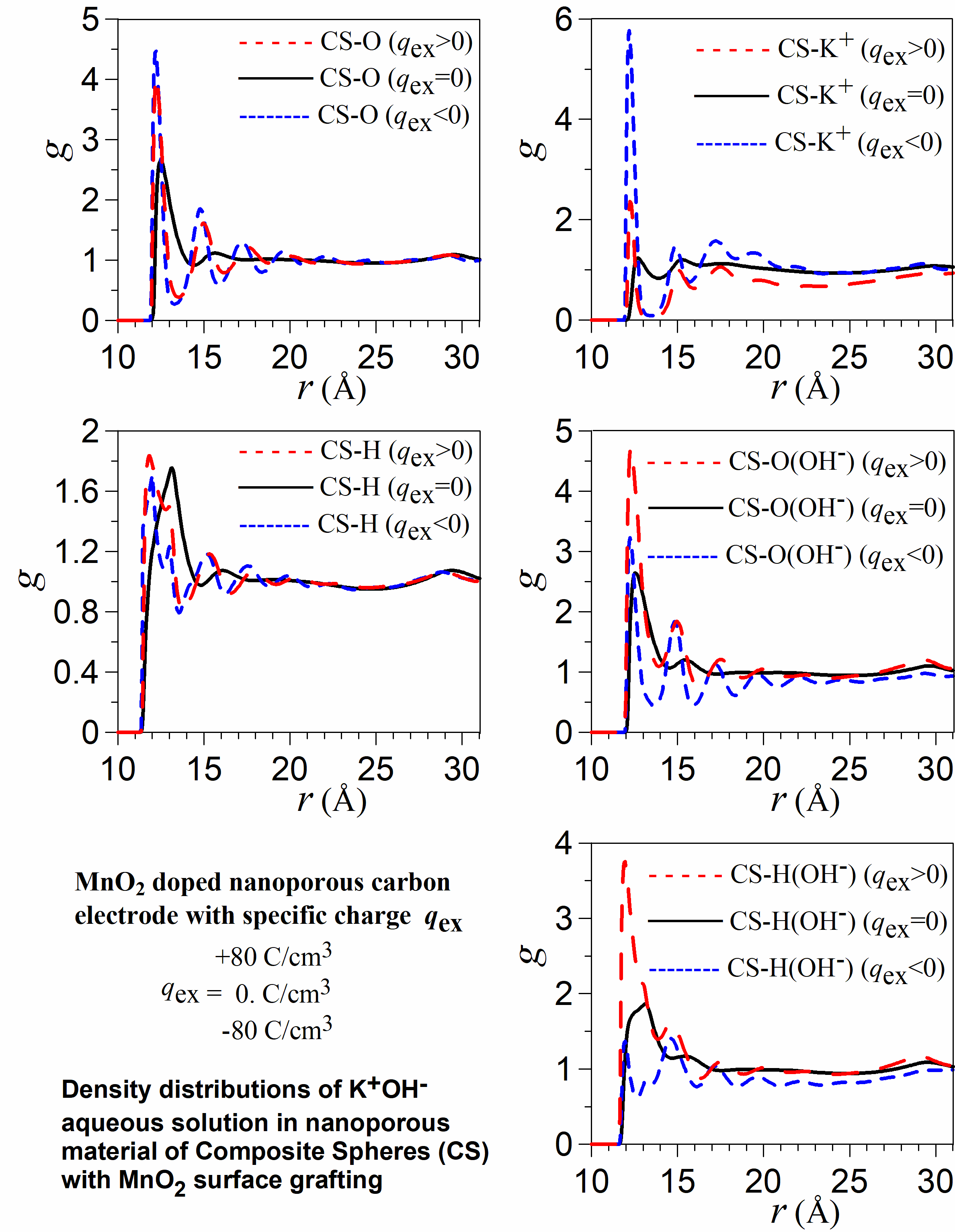}
	\caption{(Colour online) Solvation structure of KOH aqueous electrolyte solution sorbed in the nanoporous carbon electrode with the nanopores surface grafted with MnO$_2$. RDFs of water O and H sites, and of K$^+$ and OH$^-$ ions around carbon nanoparticles. Nanoporous electrode charges are the same as in figure \ref{fig:CS-KOH-g(r)}.}
	\label{fig:CS-MnO2-KOH-g(r)}
\end{figure}

\begin{figure}
	\centering
	\includegraphics[width=0.4\textwidth]{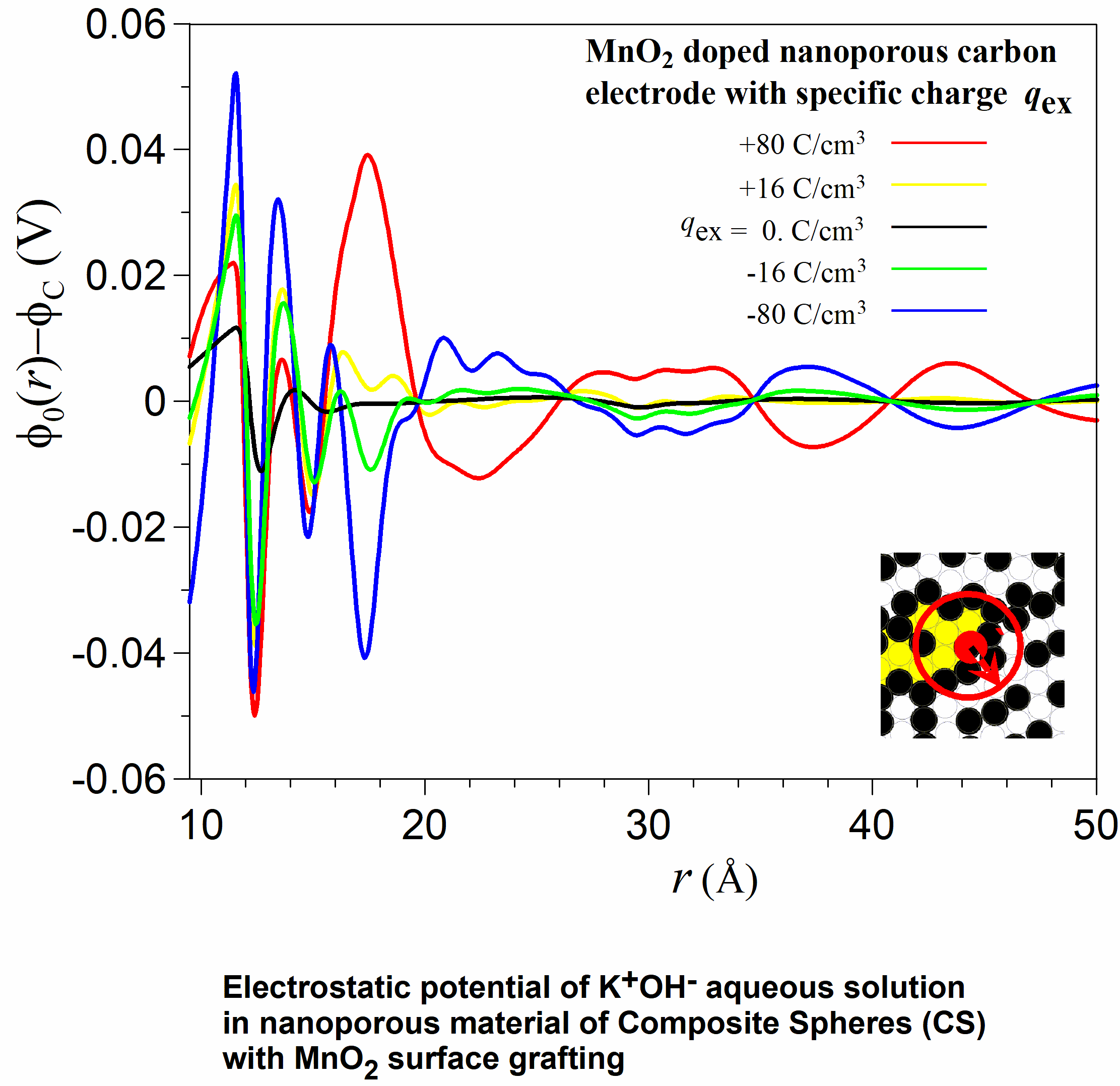}
	\caption{Electrostatic potential $\phi_0(r)$ around a nanoparticle of the nanoporous carbon electrode with respect to ``zero'' level $\phi_c$. The sorbed solution is in equilibrium with the bulk ambient aqueous solution of KOH electrolyte at concentration 120 ppm. Nanoporous electrode charges and the Inset are the same as in figure \ref{fig:CS-KOH-fi(r)}.}
	\label{fig:CS-MnO2-KOH-fi(r)}
\end{figure}

Figure \ref{fig:CS-LiKOH-g(r)} presents the RDFs of aqueous solution of the LiOH--KOH electrolyte mixture sorbed in nanoporous carbon electrodes at external charge of nanopores $q_{\rm ext}=0$ and $\pm 80$ C/cm$^3$. The behavior of K$^+$ and Li$^+$ cations as well as OH$^-$ anions is similar to a mixture of the RDFs of separate KOH and LiOH electrolytes in terms of the positions and maxima of the first three peaks of K$^+$ and Li$^+$ cations. A distinct feature is the twofold decrease of the K$^+$ maxima and some increase of the Li$^+$ maxima in the mixture, as compared to the separate solutions. On the other hand, the water distributions change and shift less for a negative electrode charge, while the distribution of water hydrogens shifts towards the nanopore, reflecting the orientation of water molecules in the mixture of KOH and LiOH electrolytes. Figure \ref{fig:CS-LiKOH-fi(r)} for the electrostatic potentials of LiKOH aqueous electrolyte solution in fact shows again a mixture of separate KOH and LiOH electrolytes. This is related to both the heights of the peaks and to the amplitudes of the oscillations.
\begin{figure}[!t]
	\centering
	\includegraphics[width=0.6\textwidth]{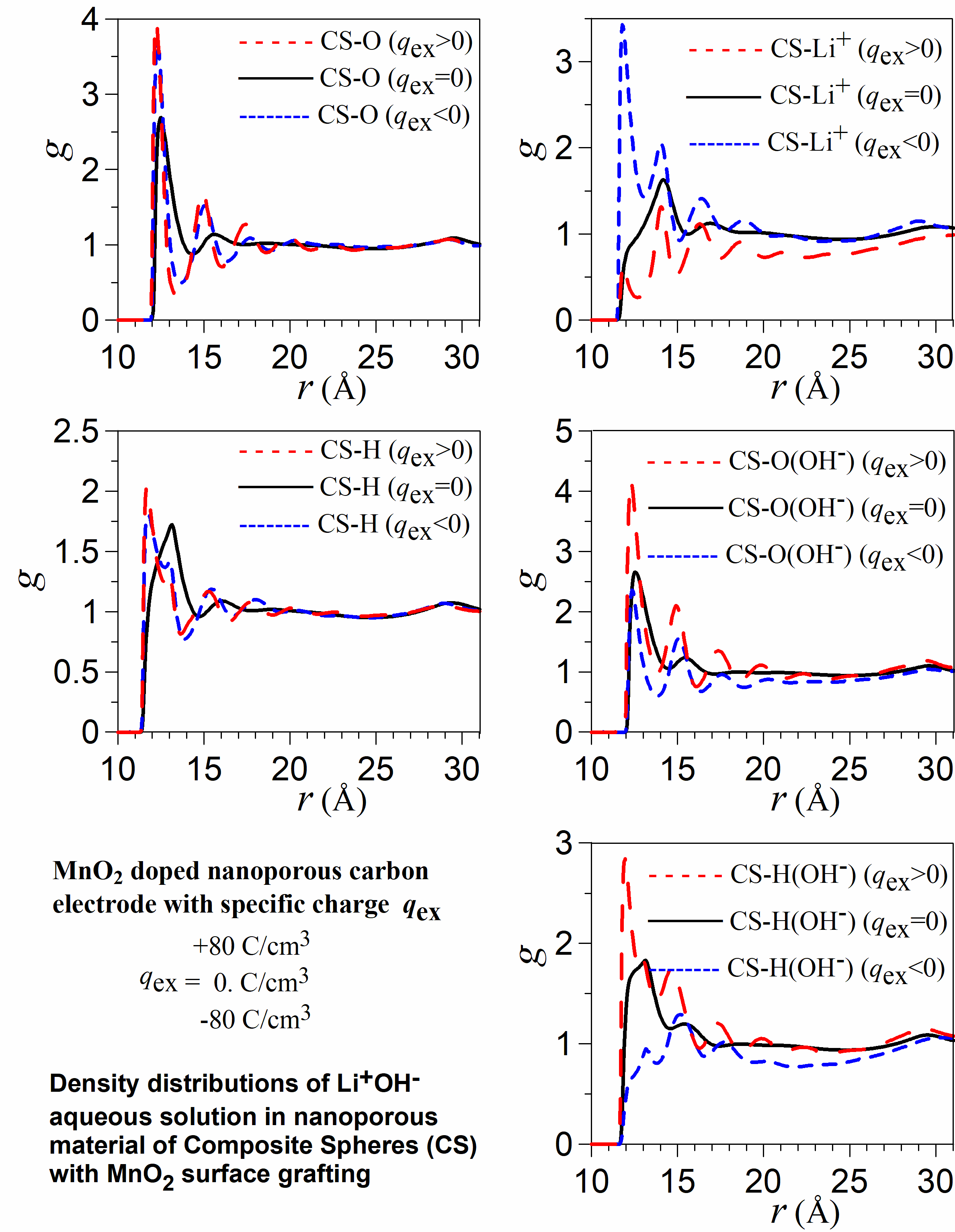}
	\caption{(Colour online) Solvation structure of LiOH aqueous electrolyte solution sorbed in the nanoporous carbon electrode with MnO$_2$ groups grafted on the inner surface of the nanopores. RDFs of water O and H sites, and of Li$^+$ and OH$^-$ ions around carbon nanoparticles. Nanoporous electrode charges are the same as in figure~\ref{fig:CS-KOH-g(r)}.}
	\label{fig:CS-MnO2-LiOH-g(r)}
\end{figure}

\begin{figure}[!t]
	\centering
	\includegraphics[width=0.4\textwidth]{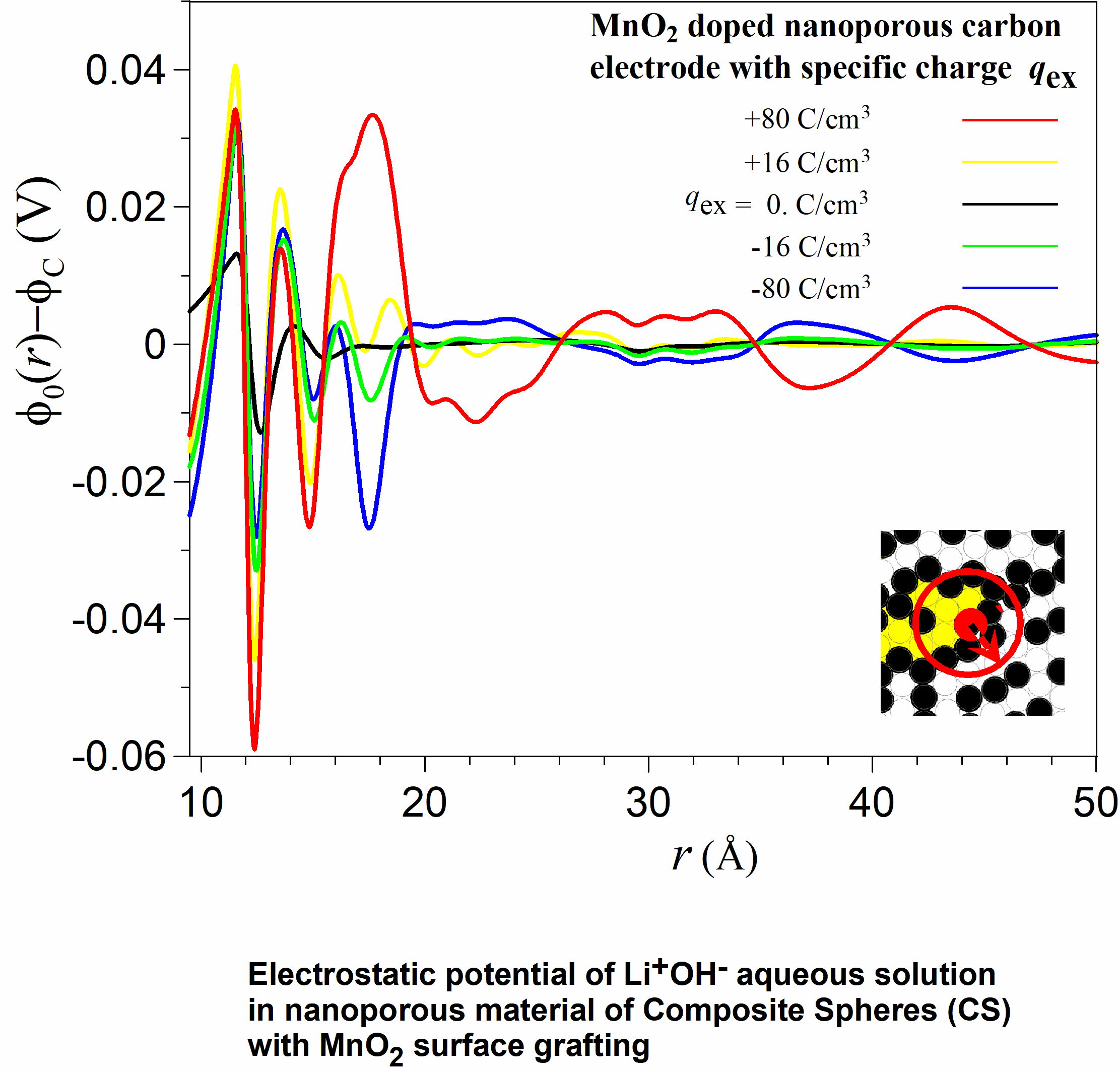}
	\caption{(Colour online) Electrostatic potential $\phi_0(r)$ around a nanoparticle of the MnO$_2$-grafted nanoporous carbon electrode with respect to ``zero'' level $\phi_c$. The sorbed solution is in equilibrium with the bulk ambient aqueous solution of LiKOH electrolyte at concentration 120 ppm. Nanoporous electrode charges and the Inset are the same as in figure \ref{fig:CS-KOH-fi(r)}.}
	\label{fig:CS-MnO2-LiOH-fi(r)}
\end{figure}

Figure \ref{fig:CS-MnO2-KOH-g(r)} depicts the changes in the RDFs of KOH aqueous electrolyte solution in the MnO$_2$-grafted nanoporous electrodes with charging from zero to specific charge $q_{\rm ext}=+80$ C/cm$^3$. Grafting the inner surface of the nanoporous electrodes with MnO$_2$ groups little changes the RDFs, except for those of K$^+$. The first peak significantly increases in both the positively and negatively charged electrode, whereas all the successive minima and peaks get lower in the MnO$_2$-grafted electrodes. The RDF of OH$^-$ oxygen is almost the same in the electrodes with and without MnO$_2$ grafting, whereas that of OH$^-$ hydrogen is significantly higher in the MnO$_2$-grafted electrode. This is related to the stronger orientational localization of OH$^-$ ions in the presence of MnO$_2$ grafting of the electrode. The electrostatic potentials of KOH electrolyte in the MnO$_2$-grafted nanoporous electrode shown in figure~\ref{fig:CS-MnO2-KOH-fi(r)} are very close to those without MnO$_2$ grafting (figure~\ref{fig:CS-KOH-fi(r)}), except for some difference in the electrode proximity where MnO$_2$ grafting molecules are located.
\begin{figure}[!t]
	\centering
	\includegraphics[width=0.55\textwidth]{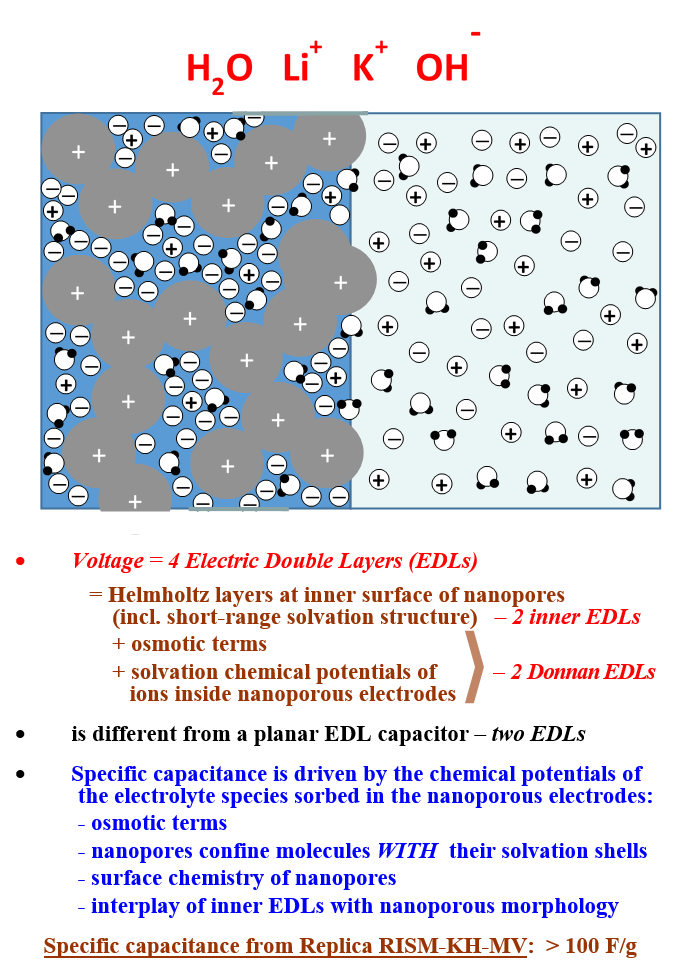}
	\caption{(Colour online) Li$^+$ and K$^+$ cations and OH$^-$ anions in aqueous solution sorbed in a nanoporous carbon electrode.}
	\label{fig:LiKOH-water-nanopores}
\end{figure}

Figure \ref{fig:CS-MnO2-LiOH-g(r)} shows the carbon-solution RDFs of LiOH electrolyte with charging the MnO$_2$-grafted nanoporous electrode from zero to specific charge $q_{\rm ext}=\pm 80$ C/cm$^3$. The MnO$_2$ grafting of the electrode significantly changes all the RDFs. At any charge of the nanoporous electrode, all the peaks and oscillations of both the water oxygen and hydrogen RDFs significantly increase in the presence of MnO$_2$ grafting. The water hydrogen RDFs also shift towards the electrode with charging, either positive or negative, reflecting the preferential orientation of water in the solvation shells of LiOH at the electrode surface. Likewise, the RDFs of Li$^+$ and OH$^-$ ions strongly change in the presence of MnO$_2$ grafting of the electrode. Namely, the heights of all the peaks significantly increase for ions with the charge sign opposite to that of the electrode, and decrease much for ions and the electrode with the same charge sign. The peaks positions do not change much with charging the electrode with either sign of the charge. The electrostatic potential of LiOH aqueous electrolyte solution in the MnO$_2$-grafted nanoporous electrode presented in figure~\ref{fig:CS-MnO2-LiOH-fi(r)} is significantly different from that without MnO$_2$ grafting. The amplitude of potential oscillations in the presence of MnO$_2$ substantially decreases and loses the long-range oscillating asymptotics observed without MnO$_2$ grafting.

Figure \ref{fig:LiKOH-water-nanopores} explains a diagram of the distributions of Li$^+$ and K$^+$ cations and OH$^-$ anions inside and at the surface of the nanoporous carbon electrode. The bulk potential level $q_{\rm ext}\phi_c$ inside the nanopores is shifted by the dipole electric field formed in the solution at the electrode macroscopic boundary due to the chemical equilibrium conditions (\ref{eqn:chemical-equilibrium-bias}). An additional EDL appearing at the macroscopic boundaries of the two electrodes counterbalances the difference between the ``interior'' part of ions in the nanoporous electrodes $k_{\rm B}T \ln\left( \rho_s \Lambda_s \right) + \Delta\mu_s$. This difference between the “interior” part of the ions in the nanoporous electrodes is counterbalanced by an additional EDL appearing at the macroscopic boundaries of the two electrodes. The chemical equilibrium between the solutions inside the nanoporous electrodes and the bulk solution outside them thus causes a major portion of the supercapacitor voltage $U(q_{\rm ext})$ \cite{Kovalenko:2017}. It appeared that the electrochemical mechanism of the nanoporous carbon supercapacitor with ambient aqueous solution of KOH electrolyte at concentration 1M is driven not just by the EDL potential change at a planar electrode but mainly by the chemical balance of the sorbed ions in the whole electrode due to the Nernst-Planck equation \cite{Taminura:2007, Kovalenko:2004, Kovalenko:2017}.

The same molecular forces determine the purification efficiency (\ref{eqn:chemical-equilibrium}) of a nanoporous electrosorption cell \cite{Kovalenko:2004}. The sorption capacity and specific capacitance are determined by an interplay of the EDL potential change in the Stern layer at the nanoparticles surface and the Gouy-Chapman layer averaged over the disordered nanoporous material, the osmotic term appearing from the difference of the ionic concentrations in the nanopores and in the electrolyte solution outside the nanoporous electrode, and the solvation chemical potentials of ions in the nanoporous material. The solvation chemical potentials of ions, solvent, and surface functional groups sorbed in the nanopores are significantly affected by their chemical specificities and steric effects. It should be noted that the enlarged effective sizes of sorbed ions affect the specific capacitance and have strong implication on supercapacitor devices. In particular, the differences of the chemical potentials in the nanopores from the solution bulk are offset by two extra EDLs at the macroscopic boundaries of the nanoporous electrodes. This major factor affects the specific capacitance and significantly changes the supercapacitor voltage.

\section{Conclusion}
The molecular mechanisms yielding high specific capacitance of supercapacitors with nanoporous electrodes and purification efficiency of nanoporous electrosorption cells are much more sophisticated than the naive scheme of a very large specific surface area of pores densely packed in nanoporous material. The mechanisms are very different from a planar electric double layer (EDL) of the equivalent surface area. This is why the specific capacitance of such a planar EDL capacitor with an insulator of ionic radii thickness would be by an order of magnitude higher than the typical values in real devices. This empirical model is typically amended by assuming the effective thickness of the EDL insulator up to 5 nm instead. That results in the empirical value of the area capacitance of 15--20~\textmu{}F/cm$^3$ for a planar EDL of equivalent area. 

As distinct, the replica RISM-KH-VM molecular theory of solvation structure, thermodynamics, and electrochemistry of electrolyte solutions sorbed in nanoporous materials reveals that the supercapacitance and sorption forces in nanoporous carbon electrodes drastically differ from a planar EDL capacitor. The chemical potentials in such systems consist of: (1) Concentration dependent ideal term of solution species in the nanoporous cathode and anode and in the bulk electrolyte solution outside these (i.e., osmotic effect); (2) Free energy of solvation in electrolyte solution sorbed in nanoporous confinement and of effective interaction with the bulk of functionalized nanoporous material, both statistically averaged over the nanoporous morphology; (3) Electric potential step of Gibbs-Donnan type that keeps the chemical balance across a diffuse double layer at the macroscopic boundary of the nanoporous electrode in contact with the bulk solution (i.e., electrode boundary effect). Further, the voltage between the cathode and anode charged carbon nanomaterials and the bulk solution outside the electrodes is determined by: (a) Electrostatic potential change from the carbon nanopores surface to the average potential level inside the nanoporous electrode, that is across the Stern surface layer and the analogue of the Gouy-Chapman diffuse layer averaged over the nanopores, plus (b) Gibbs-Donnan electric potential step from the nanoporous electrode body across its boundary to the solution bulk. 

The above picture generalizes the naive description of nanoporous electrodes based on an equivalent planar electric double layer and Donnan potential to a realistic molecular description. It is based on statistical mechanics of nanoporous systems with an interplay of electrostatics, molecular specificity of solvent and electrolyte, adsorption at functionalized surface of nanopores, accommodation of ions and their solvation shells in nanoporous confinement, and osmotic effects.

\section{Acknowledgements}

This work was supported by the Natural Sciences and Engineering Research Council of Canada (Research Grant RES0029477), Alberta Prion Research Institute Explorations VII (Research Grant RES0039402), and Alberta Innovates - Bio Solutions (Research Grant RES0023395). The computations were supported by WestGrid – Compute/Calcul Canada.

\ukrainianpart

\title{Реплічна теорія молекулярної сольватації RISM для подвійного електричного шару в нанопористих матеріалах}
\author{А. Коваленко}
\address{
	Центр програмного забезпечення для багатомасштабного моделювання,  Едмонтон, Альберта, Канада, T6E 5J5
}

\makeukrtitle

\begin{abstract}
	\tolerance=3000%
Теорія молекулярної сольватації 3D-RISM-KH застосовна в широкому спектрі задач, від енергії сольватації малих молекул до фазової поведінки полімерів та біомолекул. Вона передбачає молекулярні механізми хімічних та біомолекулярних систем. Реплічна теорія молекулярної сольватації RISM-KH-VM передбачає та пояснює структуру, термодинаміку та електрохімію розчинів електролітів, сорбованих у нанопористому матеріалі. Її перевірено на нанопористих вуглецевих суперконденсаторах з водним електролітом та нанопористих електросорбційних комірках. Фізичні механізми в цих системах обумовлюються падінням потенціалу подвійного електричного шару на шарі Штерна на поверхні нанопор та шарі Гуї-Чепмена, усередненому по нанопористому матеріалу, осмотичним доданком, зумовленим різницею іонних концентрацій у двох нанопористих електродах та в розчині електроліту зовні, а також хімічними потенціалами сольватації сорбованих іонів, усередненими по нанопористому матеріалу. Останній сильно залежить від хімічної специфіки іонів, розчинника, поверхневих функціональних груп та стеричних ефектів для сольватованих іонів, утримуваних у нанопорах.
	\keywords статистична механіка, теорія молекулярної сольватації, реплічна теорія RISM-KH-VM, електроліти, нанопористі вуглецеві суперконденсатори, електросорбційні комірки
	
\end{abstract}
\end{document}